\documentclass{article}
\usepackage{amssymb}
\usepackage{amsmath} 
\usepackage{epsfig}

\begin{document}
\numberwithin{equation}{section}

\title{Multiscale Analysis of Discrete Nonlinear Evolution Equations
\footnote{PACS \#  02-30, 03-40, \hfill Preprint \#PM 98-30}}
\author{ J. Leon, M. Manna \\
{\em Physique Math\'ematique et Th\'eorique, CNRS-UMR5825,}\\
 Universit\'e Montpellier 2, 34095 MONTPELLIER (France)}
\date{}\maketitle

\begin{abstract} The method of multiscale analysis is constructed for dicrete
systems of evolution equations for which the problem is that of the far
behavior of an input boundary datum. Discrete {\em slow space variables} are
introduced in a general setting and the related finite differences are
constructed. The method is applied to a series of representative examples: the
Toda lattice, the nonlinear Klein-Gordon chain, the Takeno system and a
discrete version of the Benjamin-Bona-Mahoney equation. Among the resulting
limit models we find a discrete nonlinear Schr\"odinger equation (with reversed
space-time), a 3-wave resonant interaction system and a discrete modified
Volterra model.  \end{abstract}

\section{Introduction}
\label{sec:intro}

The study of nonlinear dispersive waves, historically originating in water wave
problems, has received considerable attention as a consequence of two
fundamental discoveries. The first one is the concept of complete integrability
of partial differential equations discovered for the Korteveg-de Vries equation
\cite{ggkm} and soon extended to the nonlinear Schr\"odinger \cite{zakshab} and
sine-Gordon \cite{sg} equations (for a review see \cite{book1,book2,book3}). 
The second one is the modulational instability of wave trains \cite{benjamin}
which is a mechanism for the creation of localized nonlinear solitary waves,
the solitons \cite{remsnet}.

These two discoveries have a common origin, the reductive perturbation method
(or multiscale analysis) \cite{taniuti} allowing to deduce simplified equations
from a basic model without loosing its characteristic features.  The method
consists essentially in an asymptotic analysis of a perturbation series, based
on the existence of different scales.  More specifically, the method generates
a  hierarchy of (small) scales for the space and time variations of the
envelopes of a fundamental (linear) plane wave and all the overtones. The
scale is moreover directly related to the (small) amplitude of the wave
itself. The scaling of variables is performed via a Taylor expansion the
frequency $\omega(k)$ in powers of a small deviation of the wave number
$k=k_0+\epsilon\kappa$.  This deviation from the linear dispersion relation is
of course generated by the nonlinearity.

The success of the method relies mainly on the nice property that the
resulting reduced models are simple, representative and often integrable.
{\em Simple} here means actually simpler than the master equations and
allowing for usefull informations. {\em Representative} means that they
illustrate effectively real processes. This property relies on the
self consistency of the perturbation series which treats all overtones
and avoids secularities \cite{manna,mikoda}. {\em Integrable} means that
they carry an infinite set of conserved quantities, have a bi-hamiltonian
formulation, are solvable (in some sense), etc... Finally, as emphasized in
\cite{calogero},  there exists a general property of the reductive
perturbation approach which allows to understand, in a qualitative way,
why the reduced systems are {\em often} integrable.

The situation is quite different in the case of nonlinear lattices (continuous
time and discrete space) for which a reductive perturbative method does not
exist which would produce reduced {\em discrete} systems.  There are actually
three different approaches to  multiscale analysis for a discrete evolution.
The first one is obviously to go to the continuous limit right in  the starting
system, for which discreteness effect are wiped out. The second one is the {\em
semi-dicrete} appoach which consists in having a discrete carrier wave
modulated by a continuous envelope \cite{semi-dis}. In that case some
discreteness aspects are preserved, in particular, the resulting modulational
instability may depend on the carrier frequency. The third one stems from
adiabatic approximation, but the approach requires to use the {\em rotating
wave approximation} to artificially eliminate the overtones.  The price to pay
is that the predictions, e.g. the modulational instability, are not trustworthy
for large times \cite{kiv-peyr}.

We propose here a set of tools which allow to perform multiscale analysis on a
discrete evolution equation when the problem is that of the propagation of a
signal sent at one end of a nonlinear lattice.  These tools rely on the
definition of a large grid scale via the comparison of the magnitude of the
related difference operator, and on the expansion of the wave number in powers
of frequency variations due to nonlinearity.

The method will be illustrated in a series of examples (Toda lattice,
anharmonic Klein-Gordon chain, BBMP equation, ...) for which the reduced model
for the slowly varying envelope $\psi(n,t)$ results in the following evolution
(over dot stands for partial time-derivative)
\begin{equation}\label{LM}
-i\beta\left[\psi_{n+1}-\psi_{n-1}\right]+\alpha \ddot\psi_n
-\gamma|\psi_{n}|^2\psi_{n}=0\ .\end{equation}

The coefficients $\alpha$ and $\beta$ and $\gamma$ depend on the starting model
equation and on the frequency of the carrier wave. The continuous version is
well known to apply to pulse propagation in a nonlinear Kerr medium (optical
fiber) \cite{hasegawa} and has been obtained also in the context of
Rayleigh-Taylor instability and electron-beam plasma \cite{yajima} and refered
there as the {\em unstable nonlinear Schr\"odinger} equation.

Another example deals with the Takeno model of exciton-phonon coupling
in diatomic chains \cite{takeno} for which we investigate the three wave
resonant interaction. The resulting discrete system reads
\begin{align}\label{3-wave}
&\dot X+v\frac12 [X_{m+1}-X_{m-1}]=
   -\frac2{\Omega} \ a_1\bar a_2 \ , \notag\\
&\dot a_1+v_1\frac12[a_{1,m+1}-a_{1,m-1}]=
    \frac1{\nu_1}\ a_2X \ ,\\
&\dot a_2+v_2\frac12[a_{2,m+1}-a_{2,m-1}]=
-\frac1{\nu_2}\ a_1\bar X\ ,\notag
\end{align}
where $X$ is the envelope of the phonon wave (carrier frequency $\Omega$,
group velocity $v$), $a_j$ are the envelopes of the two components of the
exciton wave (carrier frequencies $\nu_j$, group velocities $v_j$), for
the Brillouin selection rule $\nu_1-\nu_2=\Omega$.

We illustrate also the method by investigating the low frequency
discrete limit of the Toda lattice \cite{toda}, and obtain
\begin{equation}\label{Langmuir}
u_{n+1}-u_{n-1}\ +\ u_n\dot u_n=0\ ,
\end{equation}
which can be viewed as the Volterra model \cite{volterra} with space and 
time exchanged.

In this paper, the emphasis will be put mainly on the method and on the above
mentionned simple examples, but not on the physical implications that one can
infer from the reduced systems. Indeed, such  studies essentially depend on the
physical context and thus require specific attention. After a short statement
of the problem in the next section, the sec. \ref{sec:diff} is devoted to the
definition of the main tools.  The sec. \ref{sec:toda} deals with the slowly
varying envelope limit of the Toda lattice, and sec. \ref{sec:nkg} with
nonlinear Klein-Gordon (or sine-Gordon) chain. The sec. \ref{sec:bbm} deals
with a new nonlinear evolution constructed as a discrete version of the 
Benjamin-Bona-Mahoney-Peregrini equation, and the sec. \ref{sec:takeno} is
devoted to the study of three wave resonant interaction in the Takeno model.

\section{Discrete waves}
\label{sec:waves}

\subsection{Boundary value problem} 

The pupose is the study of a nonlinear dispersive chain with dispersion relation
$\Omega(K)$.  The physical problem we are concerned with is the
following: the first particle of the chain (say $n=0$) is given an oscillation
(or is submitted to an external force)
at frequency $\Omega$. Would the chain be linear that this oscillation would
propagate without distorsion as the plane wave $\exp[i\Omega t+Knd)]$, with
$d$ being the lattice spacing.

But the nonlinearity induces
some deviations from the value $\Omega$, namely, the wave propagates
with actual frequency $\omega$ and wave number $k$ that we define as
\begin{equation}
\omega=\Omega+\epsilon\nu, \end{equation}
and by Taylor expansion
\begin{equation}\label{taylor-k}
k=K+\epsilon\ \frac1c\ \nu+\epsilon^2\ \gamma\ \nu^2,\end{equation}
for the following definitions (note that $c$ is the group velocity at $K$)
\begin{equation}\label{group}
\frac1c=\left.\frac{\partial k}{\partial\omega}\right|_\Omega,\quad
2\gamma=\left.\frac{\partial^2 k}{\partial\omega^2}\right|_\Omega.
\end{equation}
For notation simplicity, we shall assume everywhere that $\gamma=1$,
which does not reduce the generality of our task.

A wave packet in the {\em linear case} is then given by the Fourier
transform 
\begin{equation*}
u_{n}(t)=\int d\omega\ \hat u(\omega){\rm e}^{i(\omega t+knd)},
\end{equation*}
where $\hat u(\omega)$ has support inside the allowed band frequency 
$[\omega_0,\omega_b]$, which gives here 
$$u_n(t)=\epsilon\ {\rm e}^{i(\Omega t+Knd)}\int\ d\nu\ \hat u(\nu)
{\rm e}^{i\nu\epsilon(t+nd/c)}\ {\rm e}^{i\nu^2\epsilon^2 nd}$$
or else
\begin{equation}\label{velope}
u_n(t)= A(n,t)\psi(\xi_n,\tau_n),\quad
\left\{\begin{array}{l}
A(n,t)={\rm e}^{i(\Omega t+Knd)}\\
{\null}\\
\psi(\xi_n,\tau_n)=\epsilon\int d\nu\ \hat u(\nu)
{\rm e}^{i(\nu\tau_n+\nu^2\xi_nd)}\end{array}\right.
\end{equation}
by means of the following change of {\em independent variables}
\begin{equation}\label{change-k}
\tau_n=\epsilon(t+nd/c),\qquad \xi_n=\epsilon^2 n.\end{equation}

\subsection{Discrete scaling}
In order to keep discreteness in the space variable for the
envelope $\psi(\xi_n,\tau_n)$, we fix the small parameter as
\begin{equation}\label{eps-N}
\epsilon^2=1/N\ ,\end{equation}
and, for any given $n$, we shall consider only the
set of points $\{\cdots,\ n-N,\ n,\ n+N,\ \cdots\}$ of a {\em large grid}
indexed by the {\em slow variable} $m$, that is
\begin{equation}\label{xi-n}
\cdots\ ,\ (n-N)\to (m-1)\ ,\ n\to m\ ,\  (n+N)\to (m+1)\ ,\ \cdots
\end{equation}
To simplify the notations, we shall be using everywhere
\begin{equation}\label{nota}
\psi(\xi_{j},\tau_{j})=\tilde\psi_{j}\ ,\quad
\psi(\xi_{j},\tau_n)=\psi_{j}\ ,\end{equation}
for a given $n$ and  all $j$ (note that $\tilde\psi_n=\psi_n$). 
Hence we are interested in expressing everything in
terms of $\psi_m=\psi(m,\tau)$ defined as
\begin{equation}\label{psi-m-tau}
\psi_{n-N}=\psi_{m-1}\ ,\quad\psi_n=\psi_m\ ,
\quad\psi_{n+N}=\psi_{m+1}.\end{equation}
The problem is now to express the various difference operators occuring in
nonlinear evolutions for the product $A(n,t)\psi_m$ in terms of difference
operators for $\psi_m$.

\subsection{Initial value problem}
The traditionnal approach to multiscaling for continuous media originates
from water wave theory for which the physical problem is ususally that of
the evolution of an intial disturbance (e.g. of the surface). In that case
the observer has to follow the deformation at the (linear) group velocity.

This operation corresponds to making, in the general Fourier transform
solution, the expansion of $\omega(k)$ around small deviations of $k$ from
the linear dispersion law. The resulting change of variables (in the
discrete case for comparison) would then read
\begin{equation}\label{change-omega}
\tau_n=\epsilon^2t,\qquad \xi_n=\epsilon(nd+ct)\ ,\end{equation}
which indeed corresponds to a translation at the group velocity
$c$, that is in the {\em co-moving frame}.

In that case, the change of variable definitely breaks the discreteness of
the space variable. Hence an initial value problem for a discrete lattice
in continuous time cannot be treated within a fully dicrete multiscale
analysis.

However, when the phenomenon to observe results from a boundary input datum, the
observer stands at some given point of the lattice and compares the received
signal with the input signal. This has to be done in the {\em retarded time},
which corresponds to the change of variables \eqref{change-k}, and obviously
does  not destroy the discrete character of the variables.

\section{Difference operators}
\label{sec:diff}

\subsection{General consideration}

Before going to the technical points, it is quite usefull to recall elementary
facts concerning the relation between continuous derivation and discrete
differences.  Let $\phi_n$ denote a function of the discrete variable $n$ which
varies slowly from one site to the other.  In that case we can
{\em represent} this function by the function $f(x)=\phi_n$ of the continuous
variable $x=nd$. The constant $d$ is a {\em small arbitrary} real number
which needs not to be related to the dimension of the grid spacing 
(this would be necessary if $f(x)$ would be the datum and $\phi_n$ its
representation). By Taylor expansion we readily obtain
\begin{align}\label{def-bad}
&d\frac{\partial f(x)}{\partial x}=\phi_{n+1}-\phi_n + {\cal O}(d^2)\\
&d\frac{\partial f(x)}{\partial x}=\frac12[\phi_{n+1}-\phi_{n-1}]+
{\cal O}(d^3)\ ,\label{def-first}\\
&d^2\frac{\partial^2 f(x)}{\partial x^2}=\phi_{n+1}-2\phi_n+\phi_{n-1}+
{\cal O}(d^4)\ .\label{def-sec}
\end{align}
Consequently, if the precision of the first derivative has to exceed 
at least the value of the second derivative, it has to be defined with
\eqref{def-first}.
This is not the only reason for considering centered differences. 
The first order wave equation
\begin{equation*}
\partial_t\phi_n-v\ \frac12[\phi_{n+1}-\phi_{n-1}]=0\ ,
\end{equation*}
possess a {\em good} dispersion relation ($\omega = v\ \sin k$)
with real valued solution. Such would not be the case with the definition
\eqref{def-bad}.

Going now to the stretched space variable
$\xi=\epsilon^2 x$, we readily get
\begin{align*}
&\phi_{n\pm1}=\phi_n\pm\epsilon^2d\frac{\partial f(\xi)}{\partial \xi}+
\epsilon^4\frac12d^2\frac{\partial^2 f(\xi)}{\partial \xi^2}\pm
\epsilon^6\frac16d^3\frac{\partial^3 f(\xi)}{\partial \xi^3}+\cdots
\end{align*}
which, under discretization procedure according with rules \eqref{def-first}
\eqref{def-sec}, lead to the following {\em qualitative indications}
for the change of variable on the first and second derivatives
\begin{align}
\phi_{n+1}-\phi_{n-1}&=\epsilon^2[\phi_{m+1}-\phi_{m-1}]+
{\cal O}(\epsilon^6)\label{1-deriv-n}\\
\phi_{n+1}-2\phi_n+\phi_{n-1}&=
\epsilon^4[\phi_{m+1}-2\phi_m+\phi_{m-1}]+{\cal O}(\epsilon^8).
\label{2-deriv-n}\end{align}

\subsection{Construction of the stretched grid}

For $\epsilon^2=1/N$, the above expressions serve as a guide to the
correct rules for the change of coordinates. They can be understood simply
within the definitions \eqref{eps-N} and \eqref{xi-n} by writing
\begin{align*}
\phi_{n+N}-\phi_{n-N}&=(\phi_{n+N}-\phi_{n+N-2})+
(\phi_{n+N-2}-\phi_{n+N-4})\\
&+\cdots+(\phi_{n-N+4}-\phi_{n-N+2})+(\phi_{n-N+2}-\phi_{n-N})\\
&\sim N[\phi_{n+1}-\phi_{n-1}]
\end{align*}
where $N$ has to be chosen odd such as to find in the above expansion the
very term $\phi_{n+1}-\phi_{n-1}$. Then the meaning of this grid change
is that, on the interval $[n-N,n+N]$, the variations of $\phi_n$ are
almost equal, or else that $\phi_n$ is of almost constant slope
(with a precision of $1/N^2$).
The purpose now is to construct rigorously such simple rules, and in
particular to determine the conditions of {\em slow variation} under which 
expressions like \eqref{1-deriv-n} and  \eqref{2-deriv-n} do hold. 

To that end we define the derivatives in the original variable $n$ as
(the $C_k^\ell$ are the binomial coefficients $\frac{k!}{\ell!(k-\ell)!}$)
\begin{equation}\label{deriv-any}
\nabla \phi_n=\phi_{n+1}-\phi_{n-1}\ ,\quad
\nabla^k\phi_n=\sum_{\ell=0}^{k} (-)^\ell\ C_k^\ell\ \phi_{n+k-2\ell}\ ,
\end{equation}
and the derivatives in the new varaible $m$ defined in \eqref{xi-n} as
\begin{equation}\label{deriv-any-slow}
\Delta_N \phi_n=\phi_{n+N}-\phi_{n-N}\ ,\quad
\Delta_N^k\phi_n=\sum_{\ell=0}^{k} (-)^\ell\ C_k^\ell\ \phi_{n+N(k-2\ell)}\ .
\end{equation}

By expressing the value of a function $\phi_n$ at any other point $n+N$ in
terms of the differences $\nabla^k\phi_n$ up to $k=N$, we prove \cite{gerard}
in the appendix that, for any odd $N=2q+1$ 
\begin{align}
& \Delta_N\phi_n=\sum_{\ell=0}^{q}\alpha_q^\ell\ 
\nabla^{2\ell+1}\phi_n\ ,\quad 
& \alpha_q^\ell=\frac{(2q+1)(q+\ell)!}{(q-\ell)!(2\ell+1)!}\ ,
\label{gene-sder1}\\
& \Delta_N^2\phi_n=\sum_{\ell=1}^{2q+1}\gamma_q^\ell\ 
\nabla^{2\ell}\phi_n\ ,\quad
& \gamma_q^\ell=\frac{2(2q+1)(2q+\ell)!}{(2q+1-\ell)!(2\ell)!}\ .
\label{gene-sder2}\end{align}
The above expressions are {\em identities} which replace the analogue 
continuous change of variable $\partial_x\phi=\epsilon^2\partial_\xi\phi$.
Note that the first coeffcients are $\alpha_q^0=N$ and $\gamma_q^1=N^2$,
so as for the $p$-th order $\Delta_N$ difference whose first term is
$N^p\nabla^p\phi_n$.

\subsection{Slow variables}

We make now the hypothesis of {\em slow variation} of the function $\phi_n$ as
\begin{equation}\label{hyp-slow}
|\nabla^{k+1}\phi_n|= \epsilon^2|\nabla^k\phi_n| +{\cal O}(\epsilon^4)\ ,
\end{equation}
for some norm.
Such an hypothesis is never explicit in the continuous case where
it is usually {\em admitted implicitely} that if
$\partial_x\phi=\epsilon^2\partial_\xi\phi$, then 
$|\partial^2_x\phi|\sim \epsilon^2|\partial_x\phi|$. But actually it goes
the other way round: first one must assume that
the function $\phi(x)$ is of slow variation in $x$, i.e. 
$$\left|\frac{\partial^{k+1}\phi}{\partial x}\right|\sim 
\epsilon^2\left|\frac{\partial^{k}\phi}{\partial x}\right|\ $$
and second it is this very measure $\epsilon^2$ of the {\em velocity of 
variation} of $\phi(x)$ that allows to make the change of variable 
$\xi=\epsilon^2 x$.

It is necessary now to check that the hypothesis of slow variation is 
compatible with the expression \eqref{gene-sder1} and \eqref{gene-sder2}
as soon as $N$ is chosen equal (nearest odd integer) to $\epsilon^{-2}$.
By a recursive use of \eqref{hyp-slow} we obtain 
the following majorations at large $N=2q+1$:
\begin{align*}
&\frac1N |\Delta_N\phi_n| <  |\nabla\phi_n|
\sum_{\ell=0}^{q}\frac{\alpha_q^\ell}{(2q+1)^{2\ell+1}} \ ,\\
&\frac1{N^2}|\Delta_N^2\phi_n| <  |\nabla^2\phi_n|
\sum_{\ell=1}^{2q+1}\frac{\gamma_q^\ell}{(2q+1)^{2\ell}} \ .
\end{align*}
We compute the sums of the above series as $N=2q+1\to\infty$:
\begin{align}
&\lim_{q\to\infty}\sum_{\ell=0}^{q}\frac{\alpha_q^\ell}{(2q+1)^{2\ell+1}}=
\sum_{\ell=0}^{\infty}\frac1{2^{2\ell}(2\ell+1)!}=2 \sinh(\frac12)=1.042
\ ,\\
&\lim_{q\to\infty}\sum_{\ell=1}^{2q+1}\frac{\gamma_q^\ell}{(2q+1)^{2\ell}}=
\sum_{\ell=1}^{\infty}\frac2{(2\ell)!}=2\cosh(1)-2=1.086\ .
\end{align}
Hence the slow variation assumption are compatible with the series
\eqref{gene-sder1} and \eqref{gene-sder2}. 
The convergence is actually quite fast as, e.g. in the order 1 difference,
the first correction coefficient is $1/24$ to be compared to $2\sinh(1/2)-1$ 
(they differ by $0.5\times 10^{-3}$).

Now, with the assumption \eqref{hyp-slow}, the identities \eqref{gene-sder1}
and \eqref{gene-sder2} together with $1/N=\epsilon^2$, $N=2q+1$, lead to
\begin{align}\label{expand-nabla} 
\nabla\phi_n=\frac{1}{N}\ \Delta_N
\phi_n+{\cal O}(1/N^3)\ ,\quad \nabla^2\phi_n=\frac{1}{N^2}\ \Delta_N^2
\phi_n+{\cal O}(1/N^4)\ , \end{align} 
which are the general rules for the change of variable $n\to m$ in the first
and second differences.  These rules do concur with \eqref{1-deriv-n} and
\eqref{2-deriv-n}.

\subsection{General expressions}
Remembering the notation \eqref{nota}, by Taylor expansion in the 
continuous variable $\tau_n$, we have
\begin{align*}
&\tilde\psi_{n+1}=\psi_{n+1}+
\epsilon\frac dc\partial_\tau\psi_{n+1}+
\frac12(\epsilon\frac dc)^2\partial_\tau^2\psi_{n+1}+
\frac16(\epsilon\frac dc)^3\partial_\tau^3\psi_{n+1}\ ,\\
&\tilde\psi_{n-1}=\psi_{n-1}-
\epsilon\frac dc\partial_\tau\psi_{n-1}+
\frac12(\epsilon\frac dc)^2\partial_\tau^2\psi_{n-1}-
\frac16(\epsilon\frac dc)^3\partial_\tau^3\psi_{n-1}\ .
\end{align*}
Then we  use the identities (such general expressions are given in the
appendix)
\begin{align*}
&\psi_{n+1}-\psi_n=\frac12[\psi_{n+1}-\psi_{n-1}]+
\frac12[\psi_{n+1}-2\psi_n+\psi_{n-1}]\ ,\\
&\psi_n-\psi_{n-1}=\frac12[\psi_{n+1}-\psi_{n-1}]-
\frac12[\psi_{n+1}-2\psi_n+\psi_{n-1}]\ ,
\end{align*}
to replace hereabove $\psi_{n+1}$ and $\psi_{n-1}$ in terms of 
$\psi_n$. Finally by means of 
\eqref{expand-nabla} we arrive at the following general expressions
which constitute the basic tool allowing to compute the first and second
derivatives:
\begin{align}\label{gener-deriv+}\tilde\psi_{n+1}&=
\psi_{m}+\epsilon\frac dc\partial_\tau\psi_m+
\frac12\epsilon^2(\frac dc)^2\partial_\tau^2\psi_m+
\frac1N\frac12[\psi_{m+1}- \psi_{m-1}]\notag\\
&+\frac\epsilon N\frac dc\frac12\partial_\tau[\psi_{m+1}-\psi_{m-1}]+
\frac16\epsilon^3(\frac dc)^3\partial_\tau^3\psi_m\notag\\
&+\frac1{2N^2}[\psi_{m+1}-2\psi_{m}+\psi_{m-1}]+\frac{\epsilon^2}{4N}
(\frac dc)^2\partial_\tau^2[\psi_{m+1}- \psi_{m-1}]\notag\\
&+\frac{\epsilon^4}{24}(\frac dc)^4\partial_\tau^4\psi_m+
{\cal O}(\epsilon^5)\ ,\\
\label{gener-deriv-}\tilde\psi_{n-1}&=
\psi_{m}-\epsilon\frac dc\partial_\tau\psi_m+
\frac12\epsilon^2(\frac dc)^2\partial_\tau^2\psi_m-
\frac1N\frac12[\psi_{m+1}- \psi_{m-1}]\notag\\
&+\frac\epsilon N\frac dc\frac12\partial_\tau[\psi_{m+1}-\psi_{m-1}]-
\frac16\epsilon^3(\frac dc)^3\partial_\tau^3\psi_m\notag\\
&+\frac1{2N^2}[\psi_{m+1}-2\psi_{m}+\psi_{m-1}]-\frac{\epsilon^2}{4N}
(\frac dc)^2\partial_\tau^2[\psi_{m+1}- \psi_{m-1}]\notag\\
&+\frac{\epsilon^4}{24}(\frac dc)^4\partial_\tau^4\psi_m+
+{\cal O}(\epsilon^5).
\end{align}

Note that  pushing the general formula \eqref{expand-nabla} to higher orders,
we can reach any order of precision in such expansions.  

\subsection{Discrete derivative of a function}
By straightforward calculations, using the above basic formulae 
\eqref{gener-deriv+} and \eqref{gener-deriv-}, the first derivative follows
\begin{align}\label{first-deriv}
\tilde\psi_{n+1}-\tilde\psi_{n-1}=
2\epsilon\frac dc\partial_\tau\psi_{m}+\frac1N[\psi_{m+1}-\psi_{m-1}]
+{\cal O}(\epsilon^3)
\end{align}
together with the second derivative
\begin{align}\label{second-deriv}
\tilde\psi_{n+1}-&2\tilde\psi_{n}+\tilde\psi_{n-1}=
\epsilon^2(\frac dc)^2\partial_\tau^2\psi_{m}
+\frac\epsilon N\frac dc\ \partial_\tau[\psi_{m+1}-\psi_{m-1}]\notag\\
&+\frac1{N^2}[\psi_{m+1}-2\psi_{m}+\psi_{m-1}]
+\frac{\epsilon^4}{12}(\frac dc)^4\partial_\tau^4\psi_m
+{\cal O}(\epsilon^5).
\end{align}

It is worth remarking that, by continuation of the above formulae stopped 
at $\epsilon^3$ for the first derivative and at order $\epsilon^4$ for 
the second we obtain
\begin{align*}
&\frac{\partial}{\partial x}=\epsilon^2\ \frac{\partial}{\partial\xi}+
\frac\epsilon c\ \frac{\partial}{\partial\tau}+{\cal O}(\epsilon^3)\\
&\frac{\partial^2}{\partial x^2}=
2\frac{\epsilon^3}{c}\ \frac{\partial}{\partial\tau}
\frac{\partial}{\partial\xi}+\frac{\epsilon^2}{c^2}\ 
\frac{\partial^2}{\partial\tau^2}+{\cal O}(\epsilon^4)
\end{align*}
which are the very relations corresponding to the 
change of variable
$$\tau=\epsilon(t+x/c),\qquad \xi=\epsilon^2 x\ ,$$
continuous analogue of \eqref{change-k}.

\subsection{Discrete derivative of a product}
We have now to obtain the analogous relations for the product
$$u_n(t)= A(n,t)\psi(\xi_n,\tau_n)=A_n\tilde\psi_n$$
appearing in the definition \eqref{velope}.
The quantity $u_{n+1}-u_{n-1}$ is factorized as
\begin{align*}
&A_{n+1}\tilde\psi_{n+1}-A_{n-1}\tilde\psi_{n-1}=\\
&\frac12[A_{n+1}-A_{n-1}][\tilde\psi_{n+1}+
\tilde\psi_{n-1}]+
\frac12[A_{n+1}+A_{n-1}][\tilde\psi_{n+1}-
\tilde\psi_{n-1}].
\end{align*}
and  using the basic formula \eqref{gener-deriv+} and \eqref{gener-deriv-},
it follows, at order $\epsilon/N$ or $\epsilon^3$
\begin{align}\label{first-prod-deriv}
u_{n+1}-u_{n-1}=&[A_{n+1}-A_{n-1}]\psi_{m}+
\epsilon\frac dc[A_{n+1}+A_{n-1}]\partial_\tau\psi_{m}+\notag\\
&\frac1{N}\frac12[A_{n+1}+A_{n-1}][\psi_{m+1}-\psi_{m-1}]+\notag\\
&\epsilon^2(\frac dc)^2\frac12[A_{n+1}-A_{n-1}]
\partial_\tau^2\psi_m \ .
\end{align}
The second derivative is factorized as follows
\begin{align*}
u_{n+1}&-2u_{n}+u_{n-1}=\frac12[A_{n+1}-A_{n-1}]
[\tilde\psi_{n+1}-\tilde\psi_{n-1}]\notag\\
&+\frac12[A_{n+1}+A_{n-1}]
[\tilde\psi_{n+1}-2\tilde\psi_{n}+\tilde\psi_{n-1}]+
\tilde\psi_{n}[A_{n+1}-2A_{n}+A_{n-1}]\ , 
\end{align*}
which readily gives from \eqref{gener-deriv+}  and \eqref{gener-deriv-}
\begin{align}\label{second-prod-deriv}
u_{n+1}-2u_{n}+u_{n-1}
=&[A_{n+1}-2A_{n}+A_{n-1}]\ \psi_{m}    \notag\\
&+\epsilon [A_{n+1}-A_{n-1}]\frac dc 
\partial_\tau \psi_{m}     \notag\\
&+\epsilon^2\frac12[A_{n+1}+A_{n-1}](\frac dc)^2
\partial_\tau^2\psi_{m}             \notag\\
&+\frac1N\frac12[A_{n+1}-A_{n-1}] [\psi_{m+1}-\psi_{m-1}]\ ,
\end{align} 
at order $\epsilon/N$ or $\epsilon^3$.
The fromulae \eqref{first-prod-deriv} and \eqref{second-prod-deriv}
constitute our basic tool for deriving reduced models in the next sections.

\section{The Toda lattice}
\label{sec:toda}
The Toda chain is defined by the equation \cite{toda}
\begin{equation}\label{Toda-exp}
\ddot x_n=e^{x_{n+1}-x_n}-e^{x_n-x_{n-1}},\end{equation}
which can also be written
\begin{align}\label{toda1}
&\frac{\partial B(n,t)}{\partial t} = [1 +B(n,t)](V(n,t) -
V(n-1,t)),\\ 
\label{toda2} 
&\frac{\partial V(n,t)}{\partial t} = B(n+1,t) - B(n,t)
\end{align}
for the following definitions
\begin{equation}\label{Toda-Bn}
1+B(n,t)=e^{x_n-x_{n-1}}\ ,\quad V(n,t)=\dot x_n\ .\end{equation} 
The dispersion relation for the linearized version of the above equation
\begin{equation}\label{Toda-disp}
\Omega^2=4\sin^2\frac{Kd}2\end{equation}
is precisely that of the chain of coupled masses.

Following section \ref{sec:waves} we expand $\omega=\Omega+\epsilon\nu$ 
and consider here the two cases $\Omega=0$ and $\Omega\ne0$. The case 
$\Omega=0$ can be viewed as the low frequency limit of the Toda chain,
which is actually the standard long wave limit. The case $\Omega\ne0$
corresponds to the slowly varying envelope approximation of solutions of
the Toda lattice.

\subsection{Slowly varying envelope approximation}

We first seek a solution of equations (\ref {toda1}) and (\ref {toda2}) in the
form of a Fourier expansion in harmonics of the fundamental $A(n,t) = \exp
i(\Omega t + Knd)$ where the Fourier components are developped in a Taylor
series in power of the small parameter $\epsilon$ mesuring the amplitude of the
initial wave
\begin{align}\label{Fourier1}
&B(n,t) = \sum^{l=p}_{l=-p}\sum^{p=\infty}_{p=1}\epsilon^{p}
\psi^{(l)}_{p}(\xi_{n},\tau_{n})A^{l}(n,t),\\
\label{Fourier2}
&V(n,t) =  \sum^{l=p}_{l=-p}\sum^{p=\infty}_{p=1}\epsilon^{p}
\phi^{(l)}_{p}(\xi_{n},\tau_{n})A^{l}(n,t).
\end{align}
Note that the above series includes all overtones $A^l(n,t)=\exp il(\Omega t +
Knd)$ up to order $p$. These are generated by the nonlinear terms which
explains that the corresponding coefficients are of maximum order $\epsilon^p$.
Here we have the real-valuedness conditions 
\begin{equation}\label{real}
\psi^{(-l)}_{p} =(\psi^{(l)}_{p})^{*}\ ,\quad \phi^{(-l)}_{p} =
(\phi^{(l)}_{p})^{*}
\end{equation}
the asterisk denoting complex conjugations.
 The slow variables  $\tau_{n}$ and $\xi_{n}$ are introduced via
\begin{align}\label{varlent1}
\tau_{n} = \epsilon (t + \frac{nd}{c})\ ,\quad
\xi_{n} = \epsilon^{2}n\ ,
\end{align}
where the velocity $c$ will be determined later as a solvability
condition of equations (\ref {toda1}) and (\ref {toda2}).

By substitution of (\ref{Fourier1}) (\ref{Fourier2}) into (\ref{toda1}) and
(\ref{toda2}) and using \eqref{first-prod-deriv} (together with
$\partial_t=\epsilon\partial_\tau$) at order two in $\epsilon$ we  obtain
\begin{align}\label{eqfun1}
\sum_{l=-p}^{l=p} \sum_{p=1}^{p=\infty} &\epsilon^{p} \{ \epsilon
\frac{\partial}{\partial\tau} \psi_{p}^{(l)}(m,\tau) + i\Omega l
\psi_{p}^{(l)}(m,\tau) \} A^{l}(n,t) =\\  
&\{ 1 +  \sum^{l=p}_{l=-p}
\sum^{p=\infty}_{p=1}\epsilon^{p}\psi^{(l)}_{p}(m,\tau)A^{l}(n,t)\}
\notag\\
&\times \{\sum_{l=-p}^{l=p} \sum_{p=1}^{p=\infty} \epsilon^p [
\phi_{p}^{(l)}(m,\tau)(A^{l}(n,t) - A^{l}(n-1,t)
\notag\\
& +\epsilon(\frac{d}{c})
\frac{\partial\phi_{p}^{(l)}(m,\tau)}{\partial\tau}A^{l}(n-1,t)
-\frac{1}{2}\epsilon^{2}(\frac{d}{c})^2
\frac{\partial^{2}\phi_{p}^{(l)}(m,\tau)}{\partial\tau^{2}}A^{l}(n-1,t)
\notag\\
&+\epsilon^{2}(\phi_{p}^{(l)}(m+1,\tau) -\phi_{p}^{(l)}(m-1,\tau))
\frac{1}{2}A^{l}(n-1,t) + ...]\}  
\end{align}
\begin{align}\label{eqfun2}
\sum_{l=-p}^{l=p} \sum_{p=1}^{p=\infty} &\epsilon^{p}\{ \epsilon
\frac{\partial}{\partial\tau} \phi_{p}^{(l)}(m,\tau) + i\Omega l
\phi_{p}^{(l)}(m,\tau)\} A^{l}(n,t)
=\notag\\ 
& \sum_{l=-p}^{l=p}\epsilon^{p}\{ \psi_{p}^{(l)}(m,\tau)(A^{l}(n+1,t) 
- A^{l}(n,t))
\notag\\
&+\epsilon (\frac{d}{c}) \frac{\partial\psi_{p}^{(l)}(m,\tau)}
{\partial\tau} A^{l}(n+1,t)
\notag\\
& +\frac{1}{2} \epsilon^{2} (\frac{d}{c})^2
\frac{\partial^{2}\psi_{p}^{(l)}(m,\tau)}{\partial\tau^{2}}A^{l}(n+1,t)
\notag\\
&+\epsilon^{2}(\psi_{p}^{(l)}(m+1,\tau) -\psi_{p}^{(l)}(m-1,\tau))
\frac{1}{2}A^{l}(n+1,t) + ... \}  
\end{align}

We can proceed now to collect and solve different orders of $\epsilon^{p}$
and harmonics $l$, order $(p,l)$, in (\ref{eqfun1}) and (\ref{eqfun2}).
Note that it is enough to consider $l>0$ as negative values follow from
the reality condition \eqref{real}.
In the leading order $(1,l)$ we have
\begin{align}
&\sum_{l=-1}^{l=1} i\Omega l \psi_{1}^{(l)}(m,\tau) A^{l}(n,t) -
\sum_{l=-1}^{l=1} \phi_{1}^{(l)}(m,\tau)( A^{l}(n,t) -A^{l}(n - 1,t))= 0
\label{order1na} \\
&\sum_{l=-1}^{l=1} i\Omega l \phi_{1}^{(l)}(m,\tau) A^{l}(n,t) -
\sum_{l=-1}^{l=1} \psi_{1}^{(l)}(m,\tau)( A^{l}(n+1,t) -A^{l}(n,t))=0
\label{order1nb} 
\end{align}
This is a linear homogeneous system for
$\psi_{1}^{(l)}(m,\tau)$ and $\phi_{1}^{(l)}(m,\tau)$ polynomial in $A$.
Hence each coefficient has to vanish separately. For $l=0$ the 
system gives trivial equations. For $l=1$ the determinant of the
system for $\psi_{1}^{(1)}(m,\tau)$ and $\phi_{1}^{(1)}(m,\tau)$
is zero if $\Omega(K)$ verifies the dispersion relation
\begin{equation}
\Omega(K) = 2\sin \frac{Kd}{2}
\label{dispersion} 
\end{equation}
Under this condition we seek the general non-trivial solution
of equations (\ref{order1na}) and (\ref{order1nb})
\begin{equation}
\psi_{1}^{(1)}(m,\tau) = a\eta (m, \tau) \ ,\quad
\phi_{1}^{(1)}(m,\tau) = \eta (m, \tau)
\label{eta} 
\end{equation}
with $a$ given by
\begin{equation}
a = \exp(\frac{-iKd}{2})
\label{a} 
\end{equation}

At order $(2,l)$ we have the system
\begin{align}\label{(2,n,a)}
&\sum_{l=-2}^{l=2} \{ i\Omega l \psi_{2}^{(l)}(m,\tau) 
- \phi_{2}^{(l)}(m,\tau)(1 - a^{2l})\} A^{l}(n,t)]
 \notag\\
&+\sum_{l=-1}^{l=1}\{\frac{\partial\psi_{1}^{(l)}(m,\tau)}
{\partial\tau} 
 -(\frac{d}{c}) \frac{\partial\phi_{1}^{(l)}(m,\tau)}
{\partial\tau}a^{2l}\} A^{l}(n,t)]
\notag\\
&- \{\sum_{l=-1}^{l=1}\psi_{1}^{(l)}(m,\tau)
A^{l}(n,t) \times\sum_{l=-1}^{l=1}\phi_{1}^{(l)}(m,\tau)(1 -a^{2l})  
 A^{l}(n,t)\} = 0  
\end{align}
\begin{align}\label{(2,n,b)}
&\sum_{l=-2}^{l=2} \{i\Omega l \phi_{2}^{(l)}(m,\tau) 
- \psi_{2}^{(l)}(m,\tau)(a^{-2l} - 1)\}A^{l}(n,t)] 
\notag\\
&+\sum_{l=-1}^{l=1} \{\frac{\partial\phi_{1}^{(l)}(m,\tau)}
{\partial\tau}  
- (\frac{d}{c}) \frac{\partial\psi_{1}^{(l)}(m,\tau)}
{\partial\tau}(a^{-2l}\} A^{l}(n,t)]
= 0
\end{align}
For $l = 0$ we obtain an homogeneous system with non-zero determinant,
consequently only the trivial solution exists, so that
\begin{equation}
\psi_{1}^{(0)}(m,\tau) = 0\ ,\quad
\phi_{1}^{(0)}(m,\tau) = 0
\label{phi(1,0)} 
\end{equation}
For $l = 1$ we have an inhomogeneous linear system for $\psi_{2}^{(1)}
(m,\tau)$ and $\phi_{2}^{(1)}(m,\tau)$. The determinant of the
associated homogeneous system is zero owing to the dispersion relation
(\ref{dispersion}). Therefore, the system will have a solution if the
Fredholm solvability condition is satisfied, that is if
\begin{equation}
c = \frac{\partial \Omega(K)}{\partial K} = d\cos\frac{Kd}{2}
\label{(c)} 
\end{equation}
which determines $c$ as the group velocity.

Under this solvability condition we get
\begin{align}
&\psi_{2}^{(1)}(m,\tau) = a\delta(m,\tau) + \frac{a}{i\Omega}(a\frac{d}{c} -
1)\frac{\partial \eta(m,\tau)}{\partial\tau}
\label{psi(2,1)}\\
&\phi_{2}^{(1)}(m,\tau) = \delta(m,\tau)
\label{phi(2,1)}
\end{align}
where $\delta(m,\tau)$ is an arbitrary function. Furthermore for $l=2$ 
we obtain $\psi_{2}^{(2)}(m,\tau)$ and
 $\phi_{2}^{(2)}(m,\tau)$ 
\begin{align}
&\psi_{2}^{(2)}(m,\tau) = \frac{(a\Omega)^{2}}{2(\Omega^{2} -
\sin^{2}Kd)}\eta^{2}(m,\tau) 
\label{psi(2,2)}\\
&\phi_{2}^{(2)}(m,\tau) =\frac{ \Omega\sin(Kd)}{2(\Omega^{2} -
\sin^{2}Kd)}\eta^{2}(m,\tau) 
\label{phi(2,2)}
\end{align}

The next order (3,l) gives the system
\begin{align}\label{(3,n,a)}
&\sum_{l=-3}^{l=3} \{ i\Omega l \psi_{3}^{(l)}(m,\tau)
- \phi_{3}^{(l)}(m,\tau)(1 - a^{2l})\} A^{l}(n,t)
 \notag\\
&+\sum_{l=-2}^{l=2}\{\frac{\partial\psi_{2}^{(l)}(m,\tau)}
{\partial\tau} -(\frac{d}{c}) \frac{\partial\phi_{2}^{(l)}(m,\tau)}
{\partial\tau}a^{2l} +(\frac{1}{2})(\frac{d}{c})^{2}
\frac{\partial^{2}\phi_{1}^{(l)}(m,\tau)}{\partial\tau^{2}}a^{2l}
\notag\\
&-[ \phi_{1}^{(l)}(m +1,\tau)
-\phi_{1}^{(l)}(m-1,\tau)]\frac{a^{2l}}{2}\}A^{l}(n,t)
\notag\\
& -\sum_{l=-1}^{l=1}\psi_{1}^{(l)}(m,\tau)
A^{l}(n,t) \times \sum_{l=-2}^{l=2}\phi_{2}^{(l)}(m,\tau)(1 -a^{2l})  
 A^{l}(n,t)
\notag\\
&-\sum_{l=-2}^{l=2}\psi_{2}^{(l)}(m,\tau)
A^{l}(n,t) \times \sum_{l=-1}^{l=1}\phi_{1}^{(l)}(m,\tau)(1 -a^{2l})  
 A^{l}(n,t)
\notag\\
& -\sum_{l=-1}^{l=1}\psi_{1}^{(l)}(m,\tau)
A^{l}(n,t) \times \sum_{l=-1}^{l=1}(\frac{d}{c})\frac{\partial
\phi_{1}^{(l)}(m,\tau}{\partial\tau}a^{2l}A^{l}(n,t)
 = 0    
\end{align}
\begin{align}\label{(3,n,b)}
&\sum_{l=-3}^{l=3} \{ i\Omega l \phi_{3}^{(l)}(m,\tau)
- \psi_{3}^{(l)}(m,\tau)(a^{-2l} -1)\} A^{l}(n,t)
\notag\\
&+\sum_{l=-2}^{l=2}\{\frac{\partial\phi_{2}^{(l)}(m,\tau)}
{\partial\tau} -(\frac{d}{c}) \frac{\partial\psi_{2}^{(l)}(m,\tau)}
{\partial\tau}a^{-2l}\} A^{l}(n,t)
\notag\\
&-\sum_{l=-1}^{l=1}(\frac{1}{2})(\frac{d}{c})^{2}
\frac{\partial^{2}\psi_{1}^{(l)}(m,\tau)}{\partial\tau^{2}}a^{2l}A^{l}(n,t)
\notag\\
&-\sum_{l=-1}^{l=1}[ \psi_{1}^{(l)}(m +1,\tau)-\psi_{1}^{(l)}(m-1,\tau)]
\frac{a^{2l}}{2}A^{l}(n,t) = 0
\end{align}
The order $(3,0)$ allows us to determine $\psi_{2}^{(0)}(m,\tau)$ and
$\phi_{2}^{(0)}(m,\tau)$ via an inhomogeneous system of equations. They
read
\begin{align}
&\psi_{2}^{(0)}(m,\tau) =\frac{c^{2}}{c^{2} - d^{2}}|\eta(m,
\tau)|^{2} 
\label{psi(2,0)}\\
&\phi_{2}^{(0)}(m,\tau) =\frac{cd}{c^{2} - d^{2}}|\eta(m,\tau)|^{2} 
\label{phi(2,0)}
\end{align}

The next order $(3,1)$ is a tedious one which results in
the nonlinear evolution for $\eta(m,\tau)$. It is an inhomogeneous
linear system of equations for $\psi_{3}^{(1)}(m,\tau)$ and
$\phi_{3}^{(1)}(m,\tau)$ of determinant zero. Therefore, we will have 
a solution if the Fredholm solvability condition is satisfied. This
condition gives a nonlinear evolution of $\eta(m,\tau)$ in which
the term in $\delta(m, \tau)$ coming from  $\psi_{2}^{(1)}(m,\tau)$ and
$\phi_{2}^{(1)}(m,\tau)$ is self-eliminated. The equation for 
$\eta(m,\tau)=\eta_m$ finally reads
\begin{equation}
 -i\beta[\eta_{m+1} -\eta_{m-1}] + \alpha \frac{\partial^{2}
 \eta_m}{\partial\tau^{2}} -2|\eta_m|^{2}\eta_m= 0\ ,
\label{LM-toda}
\end{equation}
with
\begin{align}\label{param-Toda}
\alpha = \tan^{2}\frac{Kd}{2}\ ,\quad 
\beta = \sin Kd\ .
\end{align}   

\subsection{Low frequency limit}
We are interested now in the propagation of a {\em low frequency wave} and
we define the following quantities
\begin{equation}\label{Toda-def}
\omega=\epsilon\nu\ , \quad
\left.\frac{\partial\omega}{\partial k}\right|_{k=0}=v_g\equiv d,
\end{equation}
and then we develop the function $k(\omega)$ in Taylor series in
$\epsilon$. We obtain
\begin{equation}\label{Toda-k-dev}
k=\frac1d\epsilon\nu+\gamma\epsilon^3\nu^3\ ,\quad
\gamma=\frac16\ 
\left.\frac{\partial^3k}{\partial \omega^3}\right|_{k=0}.\end{equation}

From the above expansion, the new variables follow
\begin{equation}\label{Toda-var}
\tau_n=\epsilon(t+n)\ , \quad \xi_n=\epsilon^3\ n,\end{equation}
and consequently we shall consider the {\em new grid} with the definition
\begin{equation}\label{Toda-grid}
\epsilon^3=\frac1N\ ,\quad \xi_n\ \to\ m\ ,\quad \xi_{n+N}\ \to\ m+1.
\end{equation}

The method and formulae developped before applies identically except that
now we have $1/N=\epsilon^3$ instead of $\epsilon^2$.
In particular, the only formula to use is that of the
second derivative \eqref{second-deriv}.

Now we set
\begin{equation}\label{Toda-amp}
B_n(t)=\epsilon^2b(\xi_n,\tau_n),\end{equation}
and compute the limit expression of the Toda equation \eqref{Toda-Bn}.
We obtain
$$
\epsilon^4\partial_\tau\left(\frac{\partial_\tau b(m)}
{1+\epsilon^2b(m)}\right)=\epsilon^4\partial_{\tau\tau}b(m)+
\epsilon^6\partial_\tau[b(m+1)-b(m-1)]+{\cal O}(\epsilon^8),$$
which finally reduces to
\begin{equation}\label{Toda-limit}
b(m+1)-b(m-1)+b(m)\partial_\tau b(m)=0.
\end{equation}
This equation is a Volterra-like equation where one would have exchanged
space and time, as expected.

\section{Nonlinear Klein-Gordon chains}
\label{sec:nkg}

The modulation of the solutions of the Toda chain has been shown to obey the
NLS-like equation \eqref{LM} with particular values of the coefficients. It
is of interest to compare this situation resulting from an 
{\em integrable model} to the one resulting from a {\em non-integrable}
starting equation. Such is the case for the nonlinear Klein-Gordon chain
\begin{equation}\label{NKG}
\ddot u_n - \omega_1^2\left(u_{n+1}-2u_n+u_{n-1}\right)+\omega_0^2u_n
+\Gamma u_n^3=0\ ,
\end{equation}
or the sine-Gordon chain 
\begin{equation}\label{SG}
\ddot u_n - \omega_1^2\left(u_{n+1}-2u_n+u_{n-1}\right)+\omega_0^2\sin u_n
=0\  .\end{equation}
Both cases, in the perturbation scheme, are equivalent
for $\Gamma=-\omega_0^2/6$, and their dispersion relation is
\begin{equation}\label{disp-N-harm}
\Omega^2=\omega_0^2+4\omega_1^2\sin^2(\frac{Kd}2),
\end{equation}

\subsection{Evolution  of the envelope}
We start with the evolution \eqref{NKG},
and seek the evolution of $\psi_m=\psi_1^1$ with the tools developped in
sec. \ref{sec:diff} when
\begin{equation}
u(n,t)=\sum_{p=1}^{\infty}\epsilon^p\ \sum_{\ell=-p}^{\ell=p}\ 
e^{i\ell\theta(n,t)}\psi_p^{\ell}(m,\tau)
\end{equation}
with conditions on $\psi_p^{\ell}$ which ensure reality of $u_n$:
\begin{equation}
\psi_p^{\ell}=\bar\psi_p^{(-\ell)}\ ,\quad \psi_p^{(0)}\in{\mathbb R}.
\end{equation}
Here $\theta(n,t)=\Omega t+Knd$, the slow variables are those defined in
sec. \ref{sec:diff} (or those used for the Toda chain, namely
\eqref{varlent1}), and we stop everything at order $\epsilon^3=\epsilon/N$.

The coefficients of the constant term give at
order $\epsilon$ :
$$\omega_0^2\psi_1^0\quad\Rightarrow\quad\psi_1^0=0\ ,$$
atrder $\epsilon^2$ :
$$\omega_0^2\psi_2^0\quad\Rightarrow\quad\psi_1^0=0\ ,$$
and at order $\epsilon^3$ :
$$(\psi_1^0)_{\tau\tau}-\omega_1^2(\frac dc)^2(\psi_1^0)_{\tau\tau}
+\Gamma[(\psi_1^0)^3+6|\psi_1^1|^2\psi_1^0]+\omega_0^2\psi_3^0
\quad\Rightarrow\quad\psi_3^0=0\ .$$

The coefficients of $e^{i\theta}$ at
order $\epsilon$ give
$$\psi_1^1[-\Omega^2-\omega_1^2(e^{iKd}-2+e^{-iKd})+\omega_0^2]=0\ ,$$
which agrees with the dispersion relation \eqref{disp-N-harm}.
Then at order $\epsilon^2$ we obtain
\begin{align*}
&\psi_2^1[-\Omega^2-\omega_1^2(e^{iKd}-2+e^{-iKd})+\omega_0^2]+\\
&(\psi_1^1)_\tau[2i\Omega-\omega_1^2\frac dc(e^{iKd}-e^{-iKd})]=0.
\end{align*}
The first line here above vanishes due to the dispersion relation while
the second one vanishes too thanks to
\begin{equation}\label{OK}
\Omega=\omega_1^2\frac{d}c\ \sin Kd.\end{equation}
which is indeed verified as, by the definition \eqref{group}
\begin{equation}\label{celerity}
c=\frac{\partial\Omega}{\partial K}=
\frac1\Omega\ \omega_1^2d\ \sin Kd.\end{equation}

Finally the order $\epsilon^3$ leads to 
\begin{align*}
&\psi_3^1[-\Omega^2-\omega_1^2(e^{iKd}-2+e^{-iKd})+\omega_0^2]+\\
&(\psi_2^1)_\tau[2i\Omega-\omega_1^2\frac dc(e^{iKd}-e^{-iKd})]+\\
&(\psi_1^1)_{\tau\tau}[1-\omega_1^2\frac12(\frac dc)^2(e^{iKd}+e^{-iKd})]\\
&-\omega_1^2\frac12(e^{iKd}-e^{-iKd})[\psi_1^1(m+1)-\psi_1^1(m-1)]
+3\Gamma|\psi_1^1|^2\psi_1^1.
\end{align*}
The first two lines hereabove vanish identically and we are left with the
equation for $\psi=\psi_1^1$,
\begin{equation}\label{LM-NKG}
-i\beta\left[\psi_{m+1}-\psi_{m-1}\right]+\alpha \psi_{\tau\tau}(m)
+3\Gamma|\psi_{m}|^2\psi_{m}=0,\end{equation}
with the definitions
\begin{equation}\label{param-NKG}
\beta=\omega_1^2\ \sin Kd\ ,\quad
\alpha=1-\omega_1^2(\frac dc)^2\ \cos Kd.\end{equation}

The main difference between the parameter values hereabove and those for the
(integrable) Toda chain \eqref{param-Toda} is that here $\alpha$ changes
sign for some value of $K$, which is of fundamental importance for the
stability properties of the modulation. But this problem goes beyond the
scope of this paper and will be considered elsewhere.

\subsection{Continuous limit}
In order to check the consistency of our method, it is instructive to
examine the continuous limit of the nonlinear Klein-Gordon chain to verify
that its (continuous) multiscale analysis gives rise to an equation which is
precisely the continuous version of \eqref{LM-NKG}.

Defining the continuous variable $x=nd$ and the velocity $v=\omega_1d$,
the continuous limit of \eqref{NKG} reads
\begin{equation}\label{cont-NKG}
u_{tt}-v^2\ u_{xx}+\omega_0^2\ u +\Gamma u^3=0.
\end{equation}
By seeking a solution as a Fourier integral
and expanding $k$ in powers of $\epsilon$ for $\omega=\Omega+\epsilon\nu$
around $K$,
we are led as previously to the expansion
\begin{equation}\label{cont-scale}
u(x,t)=\sum_{p=1}^{\infty}\epsilon^p\ \sum_{\ell=-p}^{\ell=p}\ 
e^{i\ell\theta(x,t)}\psi_p^{\ell}(\xi,\tau)\ ,\quad \theta=\Omega t+Kx\ ,
\end{equation}
and  the change of variable ($c$ is the group velocity at frequency 
$\Omega$)
\begin{equation}
\xi=\epsilon^2\ x\ ,\qquad \tau=\epsilon\ (t+\frac xc)\ .
\end{equation}
with the  reality  conditions $\psi_p^{\ell}=\bar\psi_p^{(-\ell)}$.

Inserting everything in the evolution equation \eqref{cont-NKG} we get
as the coefficients of $e^{i\theta}$:
\begin{align*}
0=& \epsilon\left[ -\Omega^2+ v^2K^2+\omega_0^2\right]+
 \epsilon^2\left[2i\Omega \psi_\tau-2iK\frac{v^2}{c}\psi_\tau\right]+\\
& \epsilon^3\left[\psi_{\tau\tau}-2iKv^2\psi_\xi-\frac{v^2}{c^2}
\psi_{\tau\tau} +3\Gamma A|\psi|^2\psi\right] +{\cal O}(\epsilon^4)
\end{align*}
The order $\epsilon$ cancels as soon as we select the {\em linear
dispersion relation}
\begin{equation}\label{cont-disp}
\Omega^2=\omega_0^2+v^2\ K^2.\end{equation}
The order $\epsilon^2$ then {\em identically vanishes} as indeed, from the
definition of the group velocit, we  readily get
\begin{equation}\label{cont-prem}
c=v^2\ \frac K\Omega\quad \Rightarrow\quad
i\Omega-\frac{v^2}{c}\ iK=0.\end{equation}
Finally, the order $\epsilon^3$ furnishes
\begin{equation}\label{cont-NLS}
-2iKv^2\psi_\xi+[1-\frac{v^2}{c^2}]\psi_{\tau\tau}+3\Gamma|\psi|^2\psi=0.
\end{equation}
One interesting consequence here is that the coefficient of the second
derivative hereabove is from \eqref{cont-prem}
\begin{equation}\label{cont-alpha}
1-\frac{v^2}{c^2}=1-\frac1{v^2}\frac{\Omega^2}{K^2}=
-\frac{\omega_0^2}{v^2K^2}<0,\end{equation}
which never changes sign, contrarily to the discrete case.

The continuous equation \eqref{cont-NKG} being the continuous limit
of the discrete Klein-Gordon equation \eqref{NKG}, the consistency check
consists now in the verification that the continuous limit of our equation
\eqref{LM-NKG} is precisely the above NLS equation \eqref{cont-NLS}.
This readily  follows from the limits as $Kd\to0$ :
\begin{align*}
\frac12[\psi_{m+1}-\psi_{m-1}]\ &\to\ d\ \psi_\xi\\
\omega_1d                     \ &\to\  v \\
\beta=\omega_1^2\sin Kd             \ &\to\  v^2\frac Kd \\
\alpha=1-\frac{(\omega_1d)^2}{c^2}\cos Kd\ &\to\  1-\frac{v^2}{c^2}\ .
\end{align*}

\section{Discrete Benjamin-Bona-Mahoney-Peregrini\\ equation}
\label{sec:bbm}

\subsection{Construction of the model} From the quite well known Boussinesq
equations \cite{witham}, further asymptotic limits and restriction to
unidirectional propagation are possible, leading to reduced models among which
we find the Korteweg-de Vries equation (KdV) and the following
Benjamin-Bona-Mahoney-Peregrini (BBMP) equation for the field $u(x,t)$:
\begin{equation}\label{bbm1}
u_t+u_x-u_{xxt}+uu_x=0.
\end{equation}

The discrete multiscale tool, unlike in the continuous case, furnishes
expressions for the differences as infinite power series in the small
parameter. It is always more convenient to stick with the first few orders in
the expansion, like in the basic expression \eqref{first-deriv}. Consequently
we need to get rid of the third order derivative in the above evolution by
writing instead (\ref{bbm1}) as a system of equations for $u(x,t)$ and two
auxilair fields $p(x,t)$ and $v(x,t)$
\begin{align}
u_x &= p,\notag\\ 
p_t &= v,\label{bbm2}\\
v_x - u_t &=
p(1 + u).\notag
\end{align}

The discretization of a continuous model is always non unique. For instance,
the discrete analogues of KdV are the Toda lattice or the Langmuir equation,
both leading to KdV  in the continuum limit. Among different possible choices,
weintroduce here the following discrete analogue of \eqref{bbm2}:
\begin{align}
\frac{1}{2} [u(n+1,t) -u(n-1,t)] &= p(n,t),\notag\\
\frac{\partial p(n,t)}{\partial t} &= v(n,t),\label{bbm3} \\
\frac{1}{2} [v(n+1,t) -v(n-1,t)] - \frac{\partial u(n,t)}{\partial t} &=
p(n,t)[1 + u(n,t)].\notag
\end{align}

\subsection{Multiscale analysis}

As in the Toda case we seek a solution of equations \eqref{bbm3} 
in the form of a Fourier expansion in harmonics of 
the  fundamental $A(n,t) = \exp i(\Omega(K) t + Knd)$ and where the 
Fourier components are developped in a Taylor series in power of the 
small parameter $\epsilon$:
\begin{eqnarray}\label{Fbbm1}
u(n,t) &=& \sum^{l=p}_{l=-p}\sum^{p=\infty}_{p=1}\epsilon^{p}
\eta^{(l)}_{p}(\xi_{n},\tau_{n})A^{l}(n,t),\\
\label{Fbbm2}
p(n,t) &=&  \sum^{l=p}_{l=-p}\sum^{p=\infty}_{p=1}\epsilon^{p}
\psi^{(l)}_{p}(\xi_{n},\tau_{n})A^{l}(n,t),\\
\label{Fbbm3}
v(n,t) &=&  \sum^{l=p}_{l=-p}\sum^{p=\infty}_{p=1}\epsilon^{p}
\phi^{(l)}_{p}(\xi_{n},\tau_{n})A^{l}(n,t).
\end{eqnarray}
The slow variables  $\tau_{n}$ and $\xi_{n}$ are introduced via
\begin{align}\label{varlent-bbm}
\tau_{n} = \epsilon (t + \frac{nd}{c})\ ,\quad
\xi_{n} = \epsilon^{2}n\ ,
\end{align}
where the velocity $c$ will be determined later as a solvability
condition.

Using the identities 
\begin{eqnarray}\label{A1}
A^{l}(n + 1,t) + A^{l}(n - 1,t) &=& 2 \cos (lKd) A^{l}(n ,t),\\
\label{A2}
A^{l}(n + 1,t) - A^{l}(n - 1,t) &=& 2 \sin (lKd) A^{l}(n ,t),
\end{eqnarray}
the discret derivatives in \eqref{bbm3} can be written, with help of  the
derivative of a product \eqref{first-prod-deriv}, as 
\begin{align}\label{bbm8}
\frac{1}{2} (u(n+1,t) -u(n-1,t)) = &\sum^{l=p}_{l=-p}\sum^{p=\infty}_
{p=1} \epsilon^{p} \{\eta^{(l)}_{p}(m,\tau)i\sin(Kld)
\notag\\
& + \epsilon (\frac{d}{c}) 
\frac{\partial \eta^{(l)}_{p}}{\partial \tau}\cos(Kld) 
\notag\\ 
&+\frac{\epsilon^{2}}{2} [\eta^{(l)}_{p} (m + 1, \tau) - \eta^{(l)}_{p}
 (m - 1, \tau)] \cos(Kld)
\notag\\
&+\frac{\epsilon^{2}}{2}(\frac{d}{c})^{2}\frac{\partial^{2}
 \eta^{(l)}_{p}}{\partial \tau^{2}}i\sin(Kld) \}A^{l}(n,t)  
\end{align}
\begin{align}\label{bbm9}
\frac{1}{2} (v(n+1,t) -v(n-1,t)) = &\sum^{l=p}_{l=-p}\sum^{p=\infty}_
{p=1} \epsilon^{p} \{\phi^{(l)}_{p}(m,\tau)i\sin(Kld)
\notag\\
& + \epsilon (\frac{d}{c}) 
\frac{\partial \phi^{(l)}_{p}}{\partial \tau}\cos(Kld) 
\notag\\ 
&+\frac{\epsilon^{2}}{2} [\phi^{(l)}_{p} (m + 1, \tau) - \phi^{(l)}_{p}
 (m - 1, \tau)] \cos(Kld)
\notag\\
&+\frac{\epsilon^{2}}{2}(\frac{d}{c})^{2}\frac{\partial^{2}
 \phi^{(l)}_{p}}{\partial \tau^{2}}i\sin(Kld)\}A^{l}(n,t)  
\end{align}

Subtituting (\ref{bbm8}), (\ref{bbm9}) in \eqref{bbm3}, we obtain
\begin{align}\label{bbm10}
&\sum^{l=p}_{l=-p}\sum^{p=\infty}_
{p=1} \epsilon^{p} \{\eta^{(l)}_{p}(m,\tau)i\sin(Kld)
 + \epsilon (\frac{d}{c}) 
\frac{\partial \eta^{(l)}_{p}}{\partial \tau}\cos(Kld) 
\notag\\ 
&+\frac{\epsilon^{2}}{2} [\eta^{(l)}_{p} (m + 1, \tau) - \eta^{(l)}_{p}
 (m - 1, \tau)] \cos(Kld)
\notag\\
&+\frac{\epsilon^{2}}{2}(\frac{d}{c})^{2}\frac{\partial^{2}
 \eta^{(l)}_{p}}{\partial \tau^{2}}i\sin(Kld)\}A^{l}(n,t)
\notag\\
&-\sum^{l=p}_{l=-p}\sum^{p=\infty}_{p=1} \epsilon^{p} 
\psi^{(l)}_{p}(m,\tau)A^{l}(n,t) = 0.  
\end{align}
\begin{align}\label{bbm11}
&\sum^{l=p}_{l=-p}\sum^{p=\infty}_ {p=1} \epsilon^{p}\{ 
\epsilon \frac {\partial \psi^{(l)}_{p}(m,\tau)}{\partial \tau} + i\Omega l
\psi^{(l)}_{p}(m,\tau)\}A^{l}(n,t)
\notag\\
& -\sum^{l=p}_{l=-p}\sum^{p=\infty}_ {p=1} \epsilon^{p}
\phi^{(l)}_{p}(m,\tau)\}A^{l}(n,t) = 0,
\end{align}
\begin{align}\label{bbm12}
&\sum^{l=p}_{l=-p}\sum^{p=\infty}_
{p=1} \epsilon^{p} \{\phi^{(l)}_{p}(m,\tau)i\sin(Kld)
 + \epsilon (\frac{d}{c}) 
\frac{\partial \phi^{(l)}_{p}}{\partial \tau}\cos(Kld) 
\notag\\ 
&+\frac{\epsilon^{2}}{2} [\phi^{(l)}_{p} (m + 1, \tau) - \phi^{(l)}_{p}
 (m - 1, \tau)] \cos(Kld)
\notag\\
&+\frac{\epsilon^{2}}{2}(\frac{d}{c})^{2}\frac{\partial^{2}
 \phi^{(l)}_{p}}{\partial \tau^{2}}i\sin(Kld)\}A^{l}(n,t)
\notag\\
&-\sum^{l=p}_{l=-p}\sum^{p=\infty}_ {p=1} \epsilon^{p}\{ 
\epsilon \frac {\partial \eta^{(l)}_{p}(m,\tau)}{\partial \tau} + i\Omega l
\eta^{(l)}_{p}(m,\tau)\}A^{l}(n,t)
\notag\\
&- [1 + \sum^{l=p}_{l=-p}\sum^{p=\infty}_{p=1} \epsilon^{p} 
\eta^{(l)}_{p}(m,\tau) A^{l}(n,t)][\sum^{l=p}_{l=-p}\sum^{p=\infty}_{p=1} \epsilon^{p} 
\psi^{(l)}_{p}(m,\tau)A^{l}(n,t)] = 0.  
\end{align}

We proceed now to collect and solve different orders of $\epsilon^{p}$
and harmonics $l$ (order $(p,l)$) in (\ref{bbm10}) and (\ref{bbm11}) 
and (\ref{bbm12}). In the leading order $(1,l)$ we have
\begin{align}\label{1lA}
&\sum^{l=1}_{l=-1} \eta^{(l)}_{1}(m,\tau)i\sin(Kld)A^{l}(n,t)
-\sum^{l=1}_{l=-1} \psi^{(l)}_{1}(m,\tau)A^{l}(n,t) = 0,  
\\ \label{1lB}
&\sum^{l=1}_{l=-1}i\Omega l\psi^{(l)}_{1}(m,\tau)A^{l}(n,t)
-\sum^{l=1}_{l=-1} \phi^{(l)}_{1}(m,\tau)A^{l}(n,t) = 0,  
\\ \label{1lC}
&\sum^{l=1}_{l=-1} \phi^{(l)}_{1}(m,\tau)i\sin(Kld)A^{l}(n,t)
-\sum^{l=1}_{l=-1}i\Omega l\eta^{(l)}_{1}(m,\tau)A^{l}(n,t)
\notag\\
&-\sum^{l=1}_{l=-1} \psi^{(l)}_{1}(m,\tau)A^{l}(n,t) = 0. 
\end{align}
Equations (\ref{1lA}) (\ref{1lB}) and (\ref{1lC}) 
constitute a linear homogeneous system for $\eta^{(l)}_{1}$
$\psi_{1}^{(l)}$ and $\phi_{1}^{(l)}$. For $l=0$  
equations (\ref{1lA}) and (\ref{1lB}) give 
\begin{align}\label{psi10}
&\psi_{1}^{(0)}(m,\tau) = 0,
\\ \label{phi10}
&\phi_{1}^{(0)}(m,\tau) = 0,
\end{align}
and (\ref{1lC}) is satisfied. For $l=1$ the determinant of the
system for  $\eta^{(1)}_{1}$, $\psi_{1}^{(1)}$ and $\phi_{1}^{(1)}$
is zero if $\Omega(K)$ verifies the dispersion relation
\begin{equation}\label{dr}
\Omega(K) = - \frac{\sin Kd}{1 +\sin^{2} Kd }.
\end{equation}
Under this condition we arrive at  the following non-trivial solution
of equations (\ref{1lA}) (\ref{1lB}) (\ref{1lC}):
\begin{align}\label{eta11} 
&\eta^{(1)}_{1}(m,\tau) = \eta (m,\tau),
\\ \label{psi11} 
&\psi^{(1)}_{1}(m,\tau) = i\sin (Kd) \eta (m,\tau),
\\ \label{phi11} 
&\phi^{(1)}_{1}(m,\tau) = \frac{\sin^{2} Kd}{1 +\sin^{2} Kd} \eta (m,\tau),
\end{align}
where $\eta$ is now the unknown function.

At order $(2,l)$ we have the system
\begin{align}\label{2lA}
&\sum^{l=2}_{l=-2} \eta^{(l)}_{2}(m,\tau)i\sin(Kld)A^{l}(n,t)
 + \sum^{l=1}_{l=-1} (\frac{d}{c}) 
\frac{\partial \eta^{(l)}_{1}(m,\tau)}{\partial \tau}\cos(Kld)A^{l}(n,t) 
\notag\\
&-\sum^{l=2}_{l=-2} \psi^{(l)}_{2}(m,\tau)A^{l}(n,t) = 0.  
\\ \label{2lB}
&\sum^{l=1}_{l=-1} \frac {\partial \psi^{(l)}_{1}(m,\tau)}
{\partial \tau} A^{l}(n,t) + \sum^{l=2}_{l=-2} i\Omega l
\psi^{(l)}_{2}(m,\tau)A^{l}(n,t)
\notag\\
& -\sum^{l=2}_{l=-2} \phi^{(l)}_{2}(m,\tau)A^{l}(n,t) = 0,
\\ \label{2lC}
&\sum^{l=2}_{l=-2} \phi^{(l)}_{2}(m,\tau)i\sin(Kld)A^{l}(n,t)
+\sum^{l=1}_{l=-1} (\frac{d}{c}) 
\frac{\partial \phi^{(l)}_{1}}{\partial \tau}\cos(Kld)A^{l}(n,t) -
\notag\\
&\sum^{l=1}_{l=-1} \frac {\partial
\eta^{(l)}_{1}(m,\tau)}{\partial \tau}A^{l}(n,t) -
\sum^{l=2}_{l=-2}i\Omega l\eta^{(l)}_{2}(m,\tau)\}A^{l}(n,t)
-\sum^{l=2}_{l=-2}\psi^{(l)}_{2}(m,\tau)A^{l}(n,t)
\notag\\
&-\sum^{l=1}_{l=-1} \eta^{(l)}_{1}(m,\tau) A^{l}(n,t)\times
\sum^{l=1}_{l=-1} \psi^{(l)}_{1}(m,\tau)A^{l}(n,t) = 0.  
\end{align}
For $l = 0$ equation (\ref{2lB}) gives using (\ref{psi10})
\begin{align}\label{phi20}
\phi_{2}^{(0)}(m,\tau) = 0.
\end{align} 

Under (\ref {eta11}) and  (\ref {psi11}),
the equations (\ref{2lA}) and  (\ref{2lC}) constitute a homogeneous system 
for $\psi^{(2)}_{0}$ and the  $\tau $ derivative of $\eta^{(0)}_{1}$. 
The determinant is non-zero and consequently only the
trivial solution exists, so that
\begin{align}\label{psi20}
\psi^{(0)}_{2}(m,\tau) = 0,
\\ \label{eta01}
\eta^{(0)}_{1}(m,\tau) = 0.
\end{align}
For $l = 1$ we have, using (\ref{2lB}), an inhomogeneous linear 
system for $\eta_{2}^{(1)}$ and $\psi_{2}^{(1)}$. 
The determinant of the
associated homogeneous system is zero owing to the dispersion relation
(\ref{dr}). Therefore, the system will have a solution for the
Fredholm solvability condition, which is satisfied for
\begin{equation}\label{vitgroup}
c =  -d \frac{\cos^{3}(Kd)}{(1+ \sin^{2}(Kd))^{2}}=
\frac{\partial \Omega}{\partial K}\ ,
\end{equation}
which determines $c$ as the group velocity. 

It is important to rematk here that, for the Taylor expansion \eqref{taylor-k},
we must assume a nonvanishing group velocity $c$ as indeed the first term in
the expansion of $k(\omega)$ is $\epsilon/c$. Hence we must avoid here the
vicinity of the value $Kd=\pm\pi/2$ for which $c$ vanishes.  In that vicinity,
one should reconsider completely the problem.

Then for all nonvanishing values of $c$ we get
\begin{align}
&\eta_{2}^{(1)}(m,\tau) = g(m, \tau),\label{eta21}\\
&\psi_{2}^{(1)}(m,\tau) = ig(m,\tau)\sin(Kd) + (\frac{d}{c})
\frac{\partial \eta (m,\tau)}{\partial \tau} \cos(Kd),\label{psi21}\\
&\phi_{2}^{(1)}(m,\tau) = -\Omega g(m,\tau)\sin(Kd) + i
\frac{\partial \eta (m,\tau)}{\partial \tau}( \sin(Kd) -
\frac{\tan^{2}(Kd)}{\Omega}), \label{phi21}
\end{align}
where $g(m,\tau)$ is an arbitrary function. Furthermore for $l = 2$ 
we obtain $\eta_{2}^{(2)}$ and $\psi_{2}^{(2)}$ and
 $\phi_{2}^{(2)}$ as follows:
\begin{align}
&\eta_{2}^{(2)}(m,\tau) = a (Kd) \eta^{2}(m,\tau), 
       \label{eta22}\\
&\psi_{2}^{(2)}(m,\tau) = i\sin(2Kd) a(Kd)\eta^{2}(m,\tau),
       \label{psi22}\\
&\phi_{2}^{(2)}(m,\tau) = -2\Omega\sin(2Kd) a(Kd)\eta^{2}(m,\tau),
       \label{phi22}
\end{align}
where $a(Kd)$ is defined as
\begin{align}\label{A}
a(Kd) = -\frac{\sin(Kd)}{2\Omega [1+\sin^{2}(2Kd)]+\sin(2Kd)}.
\end{align}

The next order $(3,l)$ gives the system
\begin{align}\label{3lA}
&\sum^{l=3}_{l=-3} \eta^{(l)}_{3}(m,\tau)i\sin(Kld)A^{l}(n,t)
 +\sum^{l=2}_{l=-2} (\frac{d}{c}) 
\frac{\partial \eta^{(l)}_{2}}{\partial \tau}\cos(Kld)A^{l}(n,t) 
\notag\\ 
&+ \sum^{l=1}_{l=-1}\frac{1}{2} [\eta^{(l)}_{1} (m + 1, \tau) - 
\eta^{(l)}_{1}(m - 1, \tau)] \cos(Kld)A^{l}(n,t)
\notag\\
&+\sum^{l=1}_{l=-1}\frac{1}{2}(\frac{d}{c})^{2}\frac{\partial^{2}
 \eta^{(l)}_{1}}{\partial \tau^{2}}i\sin(Kld)A^{l}(n,t)
\notag\\
&-\sum^{l=3}_{l=-3}
\psi^{(l)}_{3}(m,\tau)A^{l}(n,t) = 0.  
\end{align}
\begin{align}\label{3lB}
&\sum^{l=2}_{l=-2}\frac {\partial \psi^{(l)}_{2}(m,\tau)}{\partial 
\tau}A^{l}(n,t) + \sum^{l=3}_{l=-3} i\Omega l \psi^{(l)}_{3}(m,\tau)\}A^{l}(n,t)
\notag\\
& -\sum^{l=3}_{l=-3}
\phi^{(l)}_{3}(m,\tau)\}A^{l}(n,t) = 0,
\end{align}
\begin{align}\label{3lC}
&\sum^{l=3}_{l=-3} \phi^{(l)}_{3}(m,\tau)i\sin(Kld)A^{l}(n,t)
 + \sum^{l=2}_{l=-2} (\frac{d}{c}) 
\frac{\partial \phi^{(l)}_{2}(m,\tau)}{\partial \tau}\cos(Kld)A^{l}(n,t) 
\notag\\ 
&+ \sum^{l=1}_{l=-1}\frac{1}{2} [\phi^{(l)}_{1} (m + 1, \tau) - 
\phi^{(l)}_{1}(m - 1, \tau)] \cos(Kld)A^{l}(n,t)
\notag\\
&+ \sum^{l=1}_{l=-1} \frac{1}{2}(\frac{d}{c})^{2}\frac{\partial^{2}
 \phi^{(l)}_{1}(m,\tau) }{\partial \tau^{2}}i\sin(Kld)\}A^{l}(n,t)
\notag\\
&-\sum^{l=2}_{l=-2} \frac {\partial \eta^{(l)}_{2}(m,\tau)}
{\partial \tau}A^{l}(n,t) -\sum^{l=3}_{l=-3} i\Omega l
\eta^{(l)}_{3}(m,\tau)\}A^{l}(n,t)
\notag\\
&- \sum^{l=3}_{l=-3}\psi^{(l)}_{3}(m,\tau)A^{l}(n,t)
-\sum^{l=1}_{l=-1} \eta^{(l)}_{1}(m,\tau)A^{l}(n,t)\times
\sum^{l=2}_{l=-2}\psi^{(l)}_{2}(m,\tau) A^{l}(n,t)
\notag\\
&-\sum^{l=2}_{l=-2}\eta^{(l)}_{2}(m,\tau)A^{l}(n,t)\times
\sum^{l=1}_{l=-1}\psi^{(l)}_{1}(m,\tau)A^{l}(n,t) = 0.  
\end{align}

The order $(3,0)$ determines $\eta^{(0)}_{2}$,
$\psi_{3}^{(0)}$ and $\phi_{3}^{(0)}$ via an inhomogeneous system of 
equations. They read
\begin{align}
&\eta_{2}^{(0)}(m,\tau) = -\frac{d cos(Kd)}{ d+c }|\eta(m,\tau)|^{2},
\label{eta20}\\
&\psi_{3}^{(0)}(m,\tau) = -(\frac{d^2}{c})\frac{cos(Kd)}{d +c}
|\eta(m,\tau)|^{2},\label{psi30}\\
&\phi_{3}^{(0)}(m,\tau) = 0.\label{phi30}
\end{align}

The next order $(3,1)$ allows us to find the nonlinear evolution of $\eta$. It
is an inhomogeneous linear system of equations for $\eta_{3}^{(1)}$,
$\psi_{3}^{(1)}$ and  $\phi_{3}^{(1)}$ of determinant zero. Therefore, we will
have a solution if the Fredholm solvability condition holds. This
condition gives the nonlinear evolution of $\eta$ in which the term in $g(m,
\tau)$ coming from  $\eta_{2}^{(1)}$, $\psi_{2}^{(1)}$ and $\phi_{2}^{(1)}$
cancels out. The equation for $\eta(m,\tau)=\eta_m$ reads finally
\begin{equation} \label{evol-BBMP}
-i\beta(\eta_{m+1}-\eta_{m-1})+
\alpha\frac{\partial^{2}\eta_m}{\partial^{2}\tau}-\gamma|\eta_m|^2\eta_m=0\ ,
\end{equation}
with the following definitions (we set $Kd=K$ for simplicity)
\begin{align*}
\beta=&\frac12\frac{\cos^{3}K\sin K}{1+\sin^2K}\ ,\\
\alpha=&-\frac12\ \frac{(\cos^2K-1)(\cos^2K+6)(\cos^2K-2)^2}{\cos^4K}\ ,\\
\gamma=&-\frac12\ \frac{\cos^2K-2}{\cos^2K-4}\\
&\times\frac{8\cos^6K+6\cos^5K-20\cos^4K-11\cos^3K+13\cos^2K-4}
     {4\cos^3K+3\cos^2K-\cos K+1}\ .
\end{align*}
Note that inn the Brillouin zone $K\in[-\pi,+\pi]$, the  coefficient $\alpha$
has a singularity in $K=\pm\pi/2$ where the group velocity vanishes, which is a
forbidden region.

\section{Three wave resonant interaction}
\label{sec:takeno}
To further illustrate the method,we consider here
the Takeno discrete model for interaction
of excitons (or vibrons) with the phonons in a lattice of coupled
harmonic oscillators (via a Fr\"olich-like hamiltonian)
\cite{takeno}. The model results from the Hamiltonian
\begin{align}\label{Ham-Takeno}
{\cal H}&={\cal H}_{\rm ph}+{\cal H}_{\rm ex}+{\cal H}_{\rm int}\ ,\\
&{\cal H}_{\rm ph}=
\frac12\sum_n\ M\dot u_n^2 + S[u_{n+1}-u_n]^2\notag\\
&{\cal H}_{\rm ex}=
\frac12\sum_n\ m[\dot q_n^2+\omega_0^2 q_n^2]+s[q_{n+1}-q_n]^2\notag\\
&{\cal H}_{\rm int}=
\frac12\sum_n\ A[u_{n+1}-u_{n-1}]q_n^2
\end{align}
After the rescaling
\begin{equation}
q_n=\frac1A\sqrt{2mM}q'_n\ ,\quad u_n=\frac mA u'_n
\end{equation}
and forgetting the primes, the equations of motion are
\begin{align}\label{Eq-Takeno}
&\ddot u_n - \Omega_1^2[u_{n+1}-2u_n+u_{n-1}]=
q_{n+1}^2-q_{n-1}^2\\
&\ddot q_n - \omega_1^2 [q_{n+1}-2q_n+q_{n-1}]+\omega_0^2q_n=
-q_n[u_{n+1}-u_{n-1}]\notag
\end{align}
with the following definitions
$$\Omega_1^2 =\frac SM\ ,\quad
\omega_1^2=\frac sm\ .$$

\subsection{Resonant wave interaction}
We consider now the situation of 3-wave  scattering, that is the
situation where the exciton wave contains two components (frequencies
$\nu_1$ and $\nu_2$) which interact with the phonon wave (frequency
$\Omega$) according to the Brillouin selection rule
\begin{equation}\label{brillouin}
\nu_1-\nu_2=\Omega\ ,\quad k_1-k_2=K\ .
\end{equation}
For slowly varying envelopes we set
\begin{align}
q_n(t)=&\epsilon [a_1^{(1)} e^{i\theta_1}+
a_2^{(1)} e^{i\theta_2}]\label{def-a}\\
&+\epsilon^2[a_1^{(2)} e^{2i\theta_1}+
a_2^{(2)}e^{2i\theta_2} +a_3^{(2)} e^{2i\phi}+
a_4^{(2)} e^{i(\theta_1+\theta_2)}\notag\\
&+a_5^{(2)} e^{i(\theta_1+\phi)} + a_6^{(2)} e^{i(\theta_2-\phi)}]+c.c. 
+{\cal O}(\epsilon^3) \ ,
\notag\\
u_n(t)=&\epsilon b_1^{(1)} e^{i\phi}  \label{def-b}\\
&+\epsilon^2[b_1^{(2)} e^{2i\theta_1}+
b_2^{(2)}e^{2i\theta_2} +b_3^{(2)} e^{2i\phi}+
b_4^{(2)} e^{i(\theta_1+\theta_2)}\notag\\
&+b_5^{(2)} e^{i(\theta_1+\phi)}+ b_6^{(2)} e^{i(\theta_2-\phi)}]+c.c. 
+{\cal O}(\epsilon^3) \ , \notag
\end{align}
where the amplitudes $a_i^{(j)}$ and $b_i^{(j)}$
depend on the slow variables $(m,\tau)$ and with the definitions
\begin{align}\label{var}
&\tau=\epsilon t\ ,\quad \{\xi_n=\epsilon n\, \quad \epsilon=\frac 1N\}
\to m\ ,\notag\\
&\theta_1=k_1n-\nu_1t\ ,\quad\theta_2=k_2n-\nu_2t\ ,\quad
\phi=Kn-\Omega t\ ,\\
&\theta_1-\theta_2=\phi \ .\notag\end{align} 

The choice of the above harmonics in the order $\epsilon^2$ results from
the remark that quadratic terms (like $q_nu_n$) for 3 waves induce
(with the complex conjugates) waves with phases
$2\theta_1$, $2\theta_2$, $2\phi$, $\theta_1+\theta_2$,
$\theta_1+\phi$, $\theta_2+\phi$, $\theta_1-\theta_2$,
$\theta_1-\phi$, $\theta_2-\phi.$
But the selection rules give 
$\theta_2+\phi=\theta_1$, $\theta_1-\theta_2=\phi$,
$\theta_1-\phi=\theta_2$,
which are already considered at first order and hence need not to be
included in the second order.

The above change of variables has to be applied to functions
$$\varphi_n(t)=\psi_m(\tau)e^{i(kn-\omega t)}$$
for which, using the tools developped in sec. \ref{sec:diff} now for
$N=\epsilon^{-1}$, we readily obtain at order $\epsilon$
\begin{align*}
&\ddot\varphi_n=[-\omega^2\psi_m-\epsilon 2i\omega\partial_\tau\psi_m+
\cdots] e^{i(kn-\omega t)}, \\
&\varphi_{n+1}-\varphi_{n-1}=[2i\sin k\ \psi_m+\epsilon\cos k
(\psi_{m+1}-\psi_{m-1}) +\cdots]e^{i(kn-\omega t)},\\
&\varphi_{n+1}-2\varphi_n+\varphi_{n-1}=
[2(\cos k\!-\!1)\psi_m+i\epsilon
\sin k(\psi_{m+1}-\psi_{m-1})+\cdots]e^{i(kn-\omega t)}.
\end{align*}

\subsection{The limit equation}
All the above machinery is applied now to the system \eqref{Eq-Takeno}
which gives at order $\epsilon$ :
\begin{align*}
e^{i\phi}&\ :
   \ [-\Omega^2-\Omega_1^22(\cos K-1)]b_1^{(1)}=0,\\
e^{i\theta_1}&\ :
   \ [-\nu_1^2-\omega_1^22(\cos k_1-1)+\omega_0^2]a_1^{(1)}=0,\\
e^{i\theta_2}&\ :
   \ [-\nu_2^2-\omega_1^22(\cos k_2-1)+\omega_0^2]a_2^{(1)}=0.
\end{align*}
These imply the dispersion relations
\begin{align}
\Omega^2= & 2\Omega_1^2(1-\cos K)\ ,\notag\\
\nu_1^2=\omega_0^2+2\omega_1^2(1-\cos k_1)\  &,\quad\
\nu_2^2=\omega_0^2+\omega_1^2(1-\cos k_2)\ ,
\end{align}
and hence the 3 group velocities
\begin{equation}
v=\frac{\Omega_1^2}{\Omega}\sin K\ ,\quad
v_1=\frac{\omega_1^2}{\nu_1}\sin k_1\ ,\quad
v_2=\frac{\omega_1^2}{\nu_2}\sin k_2\ .
\end{equation}

The next order, $\epsilon^2$ gives in turn
\begin{align*}
e^{i\phi}&\ :\ -2i\Omega\dot b_1^{(1)}-\Omega_1^2i\sin K
     [b^{(1)}_{1,m+1}-b^{(1)}_{1,m-1}]=4i\sin K\ a_1^{(1)}\bar a_2^{(1)},\\
e^{i\theta_1}&\ :\ -2i\nu_1\dot a_1^{(1)}-\omega_1^2i\sin k_1
     [a^{(1)}_{1,m+1}-a^{(1)}_{1,m-1}]=-2i\sin K\ a_2^{(1)}b_1^{(1)},\\
e^{i\theta_2}&\ :\ -2i\nu_2\dot a_2^{(1)}-\omega_1^2i\sin k_2
     [a^{(1)}_{2,m+1}-a^{(1)}_{2,m-1}]=2i\sin K\ a_1^{(1)}\bar b_1^{(1)},\\
e^{2i\theta_1}&\ :\ [-4\nu_1^2-2\Omega_1^2(\cos 2k_1-1)]\ b_1^{(2)}=
                    4i\sin k_1\cos k_1\ [a_1^{(1)}]^2,\\
        &\ :\ [-4\nu_1^2-2\omega_1^2(\cos 2k_1-1)+\omega_0^2]\ a_1^{(2)}=0,\\
e^{2i\theta_2}&\ :\ [-4\nu_2^2-2\Omega_1^2(\cos 2k_2-1)]\ b_2^{(2)}=
                    4i\sin k_2\cos k_2\ [a_1^{(2)}]^2,\\
        &\ :\ [-4\nu_2^2-2\omega_1^2(\cos 2k_2-1)+\omega_0^2]\ a_2^{(2)}=0,\\
e^{2i\phi}&\ :\ [-4\Omega^2-2\Omega_1^2(\cos 2K-1)]\ b_3^{(2)}=0,\\
        &\ :\ [-4\Omega^2-2\omega_1^2(\cos 2K-1)+\omega_0^2]\ a_3^{(2)}=0,\\
e^{i(\theta_1+\theta_2)}&\ :\ 
         [-4(\nu_1\!+\!\nu_2)^2-2\Omega_1^2(\cos(k_1\!+\!k_2)-1)]b_4^{(2)}=
                    4i\sin(k_1\!+\!k_2)a_1^{(1)} a_2^{(1)},\\
    &\ :\ [-4(\nu_1+\nu_2)^2-2\omega_1^2(\cos(k_1+k_2)-1)+\omega_0^2]
                                                       \ a_4^{(2)}=0,\\
e^{i(\theta_1+\phi)}&\ :\ 
         [-4(\nu_1+\phi)^2-2\Omega_1^2(\cos(k_1+K)-1)]\ b_5^{(2)}=0,\\
   &\ :\  [-4(\nu_1\!+\!\phi)^2-2\omega_1^2(\cos(k_1\!+\!K)-1)+\omega_0^2]
           a_5^{(2)}=-2i\sin K\ a_1^{(1)}b_1^{(1)},\\
e^{i(\theta_2-\phi)}&\ :\ 
         [-4(\nu_2-\phi)^2-2\Omega_1^2(\cos(k_2-K)-1)]\ b_6^{(2)}=0,\\
   &\ :\ [-4(\nu_2-\phi)^2-2\omega_1^2(\cos(k_2-K)-1)+\omega_0^2]\ a_6^{(2)}
           =2i\sin K\ a_2^{(1)}\bar b_1^{(1)}.
\end{align*}
The above first 3 equations provide the evolutions of the envelopes 
which we scale as
$$X=\sin K\ b_1^{(1)}\ , \quad a_1=\sin K\ a_1^{(1)}\ , 
\quad a_2=\sin K\ a_2^{(1)}$$
and consequently which obey
\begin{align}\label{3WRI}
&\dot X+v\frac12 [X_{m+1}-X_{m-1}]=
   -\frac2{\Omega} \ a_1\bar a_2 \ , \notag\\
&\dot a_1+v_1\frac12[a_{1,m+1}-a_{1,m-1}]=
    \frac1{\nu_1}\ a_2X \ ,\\
&\dot a_2+v_2\frac12[a_{2,m+1}-a_{2,m-1}]=
-\frac1{\nu_2}\ a_1\bar X\ .\notag
\end{align}

The remaining 12 equations give simply the coefficients of the second
harmonics in terms of those of the first, precisely:
\begin{align*}
&a_1^{(2)}=a_2^{(2)}=b_3^{(2)}=a_3^{(2)}=a_4^{(2)}=b_5^{(2)}=b_6^{(2)}=0\
,\\
&b_1^{(2)}=i\ \frac{\nu_1^2+\frac12\Omega_1^2(\cos 2k_1-1)]}
{\sin k_1\cos k_1}\ [a_1^{(1)}]^2\ ,\\
&b_2^{(2)}=i\ \frac{\nu_2^2-\frac12\Omega_1^2(\cos 2k_2-1)}
{\sin k_2\cos k_2}\ [a_1^{(2)}]^2\ ,\\
&b_4^{(2)}=i\ \frac{(\nu_1+\nu_2)^2+\frac12\Omega_1^2(\cos(k_1+k_2)-1)}
 {\sin(k_1+k_2)}\ a_1^{(1)} a_2^{(1)}\ ,\\
&a_5^{(2)}=-i\ \frac{2(\nu_1+\phi)^2+\omega_1^2(\cos(k_1+K)-1)+\omega_0^2}
{\sin K}\ a_1^{(1)}b_1^{(1)}\ ,\\
&a_6^{(2)}=i\ \frac{2(\nu_2-\phi)^2+\omega_1^2(\cos(k_2-K)-1)+\omega_0^2}
{\sin K}\ a_2^{(1)}\bar b_1^{(1)}\ .
\end{align*}
Hence eqs. \eqref{3WRI} constitute a closed system of equations for the
first order variations of the envelopes. It is the discrete analogue of the
continuous 3-wave resonant interaction system.

\appendix
\section{Discrete change of variable}

 For a given function $x$ of the discrete variable $n$, we are intrested in
expressing all-order differences in a {\em large grid} indexed by the variable
$m$ (taking points separated by a given odd interger $p$) in terms of the
original hierarchy of differences \eqref{deriv-any}.  The following
notations will be used throughout
\begin{align}\label{deriv-any-bis} 
\nabla^\ell x_n  &=  \sum_{k=0}^{\ell} (-)^k\
C_\ell^k\ x_{n+\ell-2k}=x_n^{(\ell)}\ , \\
\Delta_p^\ell x_n  &=  \sum_{k=0}^{\ell} (-)^k\
C_\ell^k\ x_{n+p(\ell-2k)}=x_m^{(\ell)}\ .\label{deriv-any-slow-bis}
\end{align} 
To express the $\Delta$-differences in terms of the $\nabla$-differences, 
we first write {\em exact} Taylor like series as the following
{\em identities} (valid for any given $n$ and any point function $x_n$) 
\begin{equation}\label{taylor-gene} 
x_{n+p}=\sum_{j=0}^p a_p^jx_n^{(j)}\ +\ \frac12\tilde x_n^{(p)}\ ,\quad 
x_{n-p}=\sum_{j=0}^p(-)^ja_p^jx_n^{(j)}\ +(-)^{p+1}\frac12\tilde x_n^{(p)} \ ,
\end{equation} 
with the following definitions for the {\em tilde}-derivatives 
\begin{equation}\label{tilde-deriv}
\tilde x_n^{(p)}=x_{n+p}+ \sum_{\ell=1}^{p} (-)^\ell\ C_{p+1}^\ell\
x_{n+p+1-2\ell}+(-)^{p+1}x_{n-p}\ .  \end{equation}  
For instance we may write
\begin{align*} x_{n+5}=& x_{n}+\frac52
x_n^{(1)}+\frac92 x_n^{(2)}+\frac52 x_n^{(3)} +\frac62 x_n^{(4)}+\frac12
x_n^{(5)}+\frac12 \tilde x_n^{(5)}\\ x_{n+4}=& x_{n}+\frac52 x_n^{(1)}+\frac42
x_n^{(2)}+\frac52 x_n^{(3)} +\frac12 x_n^{(4)}+\frac12 \tilde x_n^{(4)}\\
x_{n+3}=& x_{n}+\frac32 x_n^{(1)}+\frac42 x_n^{(2)}+\frac12 x_n^{(3)} +\frac12
\tilde x_n^{(3)}\\ x_{n+2}=& x_{n}+\frac32 x_n^{(1)}+\frac12 x_n^{(2)} +\frac12
\tilde x_n^{(2)}\\ x_{n+1}=& x_{n}+\frac12 x_n^{(1)}+\frac12 \tilde x_n^{(1)}\\
x_{n-1}=& x_{n}-\frac12 x_n^{(1)}+\frac12 \tilde x_n^{(1)}\\ x_{n-2}=&
x_{n}-\frac32 x_n^{(1)}+\frac12 x_n^{(2)} -\frac12 \tilde x_n^{(2)}\\ x_{n-3}=&
x_{n}-\frac32 x_n^{(1)}+\frac42 x_n^{(2)}-\frac12 x_n^{(3)} +\frac12 \tilde
x_n^{(3)}\\ x_{n-4}=& x_{n}-\frac52 x_n^{(1)}+\frac42 x_n^{(2)}-\frac52
x_n^{(3)} +\frac12 x_n^{(4)}-\frac12 \tilde x_n^{(4)}\\ x_{n-5}=& x_{n}-\frac52
x_n^{(1)}+\frac92 x_n^{(2)}-\frac52 x_n^{(3)} +\frac62 x_n^{(4)}-\frac12
x_n^{(5)}+\frac12 \tilde x_n^{(5)}\\ 
\end{align*} 
which are useful to play with in order to understand in a concrete way the
relations that follow.

The $a_p^j$ are the coefficients to be computed but only some of them are
needed, as indeed we consider {\em differences} $\Delta_p^k x_n$.
 For the definitions
\begin{equation}
\alpha_q^\ell=2a_{2q+1}^{2\ell +1}\ ,\quad
\beta_r^\ell=2a_{2r}^{2\ell}\ ,
\end{equation}
the first and second derivatives read ($p=2q+1$) 
\begin{align} \label{ident-diff}
&x_{n+p}-x_{n-p}=\sum_{\ell=0}^{q}\alpha_q^\ell\ \nabla^{2\ell+1}x_n\ ,\\
&x_{n+2r}-2x_n+x_{n-2r}=\sum_{\ell=1}^{r}\beta_r^\ell\ \nabla^{2\ell}x_n\ ,
\label{ident-diff-bis}
\end{align}
and we prove hereafter the expressions \eqref{gene-sder1} and
\eqref{gene-sder2} for the coefficients $\alpha_q^\ell$ and $\beta_r^\ell$
\cite{gerard}:
\begin{equation}\label{alpha-beta}
\alpha_q^\ell=\frac{(2q+1)(q+\ell)!}{(q-\ell)!(2\ell+1)!}\ ,
\quad   \beta_r^\ell=\frac{2r(r+\ell-1)!}{(r-\ell)!(2\ell)!}\ .
\end{equation}
Note that in \eqref{gene-sder2} we simply have set $\gamma_q^\ell=\beta_r^\ell$
for $r=2q+1$.

As the point $n$ is fixed (arbitrary) lighter notations can be used, namely
\begin{align}\label{def-f}
& G_q=x_n^{(2q+1)} \ ,\quad &  A_q=x_{n+(2q+1)}-x_{n-(2q+1)}\ ,\notag\\
& H_r=x_n^{(2r)}\ ,\quad & S_0=x_n\ ,\quad S_r=x_{n+2r}+x_{n-2r}\ ,
\end{align}
and the definition \eqref{deriv-any-bis} can now be written
\begin{equation}\label{deriv-odd}
G_q=\sum_{k=0}^q (-)^{q-k}\ C_{2q+1}^{q-k}\ A_k\ ,\quad 
H_r=\sum_{k=0}^r (-)^{r-k}\ C_{2r}^{r-k}\ S_k\ ,
\end{equation}
while \eqref{ident-diff} and \eqref{ident-diff-bis} become respectively
(note that $\beta_r^0=2$)
\begin{equation}\label{id-di}
A_q=\sum_{\ell=0}^q\alpha_q^\ell\ G_\ell\ ,\quad
S_r=\sum_{\ell=0}^r\beta_r^\ell\ H_r\ .
\end{equation}
Note that neither in $A_q$ nor in $S_r$ appear the tilde-differences
\eqref{tilde-deriv}.

The first step is the computation of the coefficients $\alpha_q^\ell$.
By replacing \eqref{deriv-odd} in \eqref{id-di}, we arrive at
the following equivalent relations
\begin{align} \label{ident-diff-2}
&A_q=\sum_{\ell=0}^{q}\alpha_q^\ell\ 
\sum_{k=0}^\ell (-)^{\ell-k}\ C_{2\ell+1}^{\ell-k}\ A_k\ ,\\
 \label{ident-diff-3}
&G_q=\sum_{k=0}^q (-)^{q-k}\ C_{2q+1}^{q-k}\ 
\sum_{\ell=0}^{k}\alpha_k^\ell G_\ell\ ,\end{align}
which are {\em identities} valid for any function $x_n$, that is for any
choice of the sequences $\{ A_k\}$ and $\{G_\ell\}$.
Consequently the coefficients of each $A_k$ and $G_k$ identically vanish,
which furnishes the equivalent recursion relations 
\begin{align}
\label{alpha-q} 
&\forall\{\ell,q\} \ ,\quad \ell<q\ ,\quad
\sum_{k=\ell}^q (-)^{k-\ell}\ C_{2k+1}^{k-\ell}\ \alpha_q^k=0
 \ ,\quad\alpha_q^q=1\ ,\\
\label{alpha-ell}
&\forall\{\ell,q\} \ ,\quad \ell<q\ ,\quad
\sum_{k=\ell}^q (-)^{q-k}\ C_{2q+1}^{q-k}\ \alpha_k^\ell=0
 \ ,\quad\alpha_q^q=1\ .
\end{align}

The rest of the proof is the check that the expression of $\alpha_q^k$ given in
\eqref{alpha-beta} does solve the above recursion relations. 
This is easily done with help of the following identity
\begin{equation}
\sum_{m=0}^n (-)^m\frac{(s+n+m-1)!}{m!(n-m)!(s+m)!}=0
\end{equation}
which can be obtained by differientiating $n$ times a conveniently chosen
polynomial of degree $n-1$ in the variable $s$, namely
\begin{align}
&Df(s)=f(s+1)-f(s)\ ,\quad D^nf(s)=\sum_{m=0}^n (-)^{n-m}C_n^m\ f(s+m)\ ,
\label{non-center}\\
&f(s)=\frac{(s+n-2)!}{(s-1)!}\quad\Rightarrow\quad D^nf(s)=0\ .\label{gerard}
\end{align}
The same procedure is then applied to get the coefficients $\beta_q^k$
as given in \eqref{alpha-beta}. 

{\bf Aknowledgements:} It is a pleasure to aknowledge many helpful discussions
with G. Mennessier whose contribution has been essential in many technical
points.


\end{document}